\UseRawInputEncoding

\documentclass[aps,pra,reprint,amssymb,superscriptaddress]{revtex4-2} 
\usepackage{amsmath,amssymb,amsfonts}
\usepackage{graphicx}
\usepackage{dcolumn}
\usepackage{bm}
\usepackage{booktabs}
\usepackage{subcaption}
\usepackage{multirow}
\usepackage[colorlinks=true,citecolor=blue,linkcolor=blue,urlcolor=blue]{hyperref}
\usepackage[mathlines]{lineno}
\usepackage{orcidlink}

\usepackage{float}
\makeatletter
\let\newfloat\newfloat@ltx
\makeatother
\usepackage{algorithm}
\usepackage{algpseudocode}

\usepackage{braket}
\usepackage[dvipsnames]{xcolor}

\begin{document}

\title{Resource assessment of classical and quantum hardware for post-quench dynamics}

\author{Joseph Vovrosh\,\orcidlink{0000-0002-1799-2830}}
\email{joseph.vovrosh@pasqal.com}
\let\comma,
\affiliation{PASQAL SAS, 24 rue Emile Baudot - 91120 Palaiseau,  Paris, France}

\author{Tiago Mendes-Santos\,\orcidlink{0000-0001-6827-5260}}
\let\comma,
\affiliation{PASQAL SAS, 24 rue Emile Baudot - 91120 Palaiseau,  Paris, France}

\author{Hadriel Mamann\,\orcidlink{0009-0002-3832-6471}}
\let\comma,
\affiliation{PASQAL SAS, 24 rue Emile Baudot - 91120 Palaiseau,  Paris, France}

\author{Kemal Bidzhiev}
\let\comma,
\affiliation{PASQAL SAS, 24 rue Emile Baudot - 91120 Palaiseau,  Paris, France}

\author{Fergus Hayes\,\orcidlink{0000-0001-7628-3826}}
\let\comma,
\affiliation{PASQAL SAS, 24 rue Emile Baudot - 91120 Palaiseau,  Paris, France}

\author{Bruno Ximenez\,\orcidlink{0009-0006-8985-1355}}
\let\comma,
\affiliation{PASQAL SAS, 24 rue Emile Baudot - 91120 Palaiseau,  Paris, France}

\author{Lucas B\'eguin\,\orcidlink{0000-0003-1388-0791}}
\let\comma,
\affiliation{PASQAL SAS, 24 rue Emile Baudot - 91120 Palaiseau,  Paris, France}

\author{Constantin Dalyac\,\orcidlink{0000-0002-0339-6421}}
\let\comma,
\affiliation{PASQAL SAS, 24 rue Emile Baudot - 91120 Palaiseau,  Paris, France}

\author{Alexandre Dauphin\,\orcidlink{0000-0003-4996-2561}}
\email{alexandre.dauphin@pasqal.com}
\let\comma,
\affiliation{PASQAL SAS, 24 rue Emile Baudot - 91120 Palaiseau,  Paris, France}

\begin{abstract}
We estimate the run-time and energy consumption of simulating non-equilibrium dynamics on neutral atom quantum computers in analog mode, directly comparing their performance to state-of-the-art classical methods, namely Matrix Product States and Neural Quantum States. By collecting both experimental data from a quantum processing unit (QPU) in analog mode and numerical benchmarks, we enable accurate predictions of run-time and energy consumption for large-scale simulations on both QPUs and classical systems through fitting of theoretical scaling laws. Our analysis shows that neutral atom devices are already operating in a competitive regime, achieving comparable or superior performance to classical approaches while consuming significantly less energy. These results demonstrate the potential of analog neutral atom quantum computing for energy-efficient simulation and highlight a viable path toward sustainable computational strategies.
\end{abstract}

\date{\today}

\maketitle

\section{\label{sec:intro}Introduction}
The development in the late 80s of new algorithms tailored for quantum computers marked the beginning of a new interdisciplinary field at the intersection of quantum physics, information theory, and computer science. Quantum computing’s original promise, grounded in complexity theory, stemmed from the discovery of algorithms offering exponential speedups over classical methods, and from evidence that quantum computers may be able to solve problems that are intractable for nondeterministic classical algorithms~\cite{bernstein1993quantum,nielsen2010quantum}. With recent technological advances demonstrating key building blocks for fault-tolerant quantum computers, more systematic resource estimation have been performed for various applications to evaluate their utility~\cite{fellous2021limitations,van2023using,gidney2025factor,harrigan2024expressing,nguyen2024quantum,agrawal2024quantifying,bellonzi2024feasibility}. 

These results, however, describe asymptotic regimes for fault-tolerant quantum computers. In contrast, current quantum processors operate at the scale of hundreds of qubits, with a limited circuit depth, and with noise, far from the asymptotic limit where such separations are expected to appear. It is therefore timely to evaluate today's devices through resource-based metrics, such as run-time efficiency and energetic cost on specific problems. Beyond computational metrics, energy consumption has become an increasingly important dimension of evaluation~\cite{auffeves2022quantum,jaschke2023quantum}. This is particularly relevant given that classical methods, such as large-scale AI or high performance scientific computing, are demanding ever-growing amounts of energy~\cite{verdecchia2023systematicreviewgreenai, schwartz2019greenai,suarez2025energy}. As computing contributes a growing share of global energy consumption, energy-to-solution benchmarks for quantum processors are needed to evaluate whether they can provide not only computational advantages but also more sustainable pathways to scaling~\cite{jaschke2023quantum}.

\begin{figure}[!h]
    \includegraphics[width=0.40\textwidth]{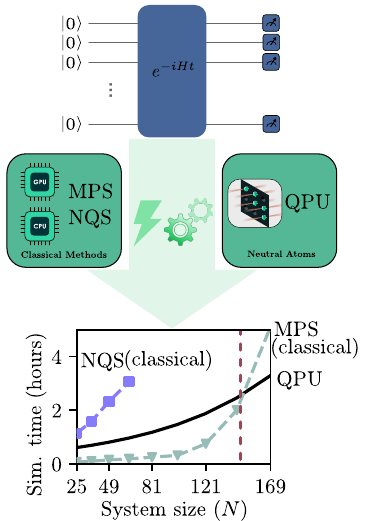}
    \caption{Resource estimation for simulating post-quench dynamics of an Ising Hamiltonian, showing both computational time and energy consumption on classical and quantum hardware. The classical data is measured from actual simulations, while the QPU data is estimated based on performance data. The figure highlights the crossover point where classical simulations become less efficient and quantum hardware is expected to offer an advantage.}
    \label{fig:Fig_1}
\end{figure}

This reframing shifts the focus toward identifying concrete regimes in which existing hardware can already demonstrate practical benefits over classical approaches. This strategy is also adopted in two recent independent works~\cite{lanes2025frameworkquantumadvantage,huang2025vast} proposing definitions of quantum advantage. In particular, the authors use concepts such as quantum utility~\cite{lanes2025frameworkquantumadvantage} or empirical quantum advantage~\cite{huang2025vast}.  For example, a 100x speedup in run-time with respect to classical computers with the same accuracy means that computations that currently take months on a classical computer could be done in a day on a quantum computer. Given a limited time budget, this speedup would allow a broader scan of the parameter space.

While such benchmarks have been developed for superconducting~\cite{arute2019quantum, Morvan_2024}, trapped-ions{~\cite{haghshenas2025digitalquantummagnetismfrontier, góis2024energeticquantumadvantagetrappedion} and photonic platforms~\cite{liu2025robustquantumcomputationaladvantage,cimini2020experimental}, neutral-atom Quantum Processing Units (QPU) remain comparatively not yet much explored~\cite{zhou2025resourceanalysislowoverheadtransversal,choi2023preparing,shaw2024benchmarking}. This gap is especially relevant for current devices operated in analog mode~\cite{Henriet_2020}, where precisely controlled Rydberg interactions among trapped atoms provide a natural substrate for quantum simulation~\cite{Browaeys_2020}. These systems have already enabled advances in probing ground-state physics~\cite{Scholl_2021, Semeghini_2021,ebadi2021quantum,zhang2025probing}, exploring non-equilibrium dynamics~\cite{Bernien_2017,bluvstein2021controlling,manovitz2025quantum, vovrosh2025mesonspectroscopyexoticsymmetries,gonzalez2025observation}, and tackling graph-based problems~\cite{dalyac2024graph}, including optimization~\cite{Ebadi_2022, cazals2025identifyinghardnativeinstances} and machine learning~\cite{ henry2021quantum,cong2019quantum,kornjavca2024large} applications.

In this work, we address this gap by analyzing neutral-atom analog QPUs through the lens of resource-based benchmarking. Rather than contrasting classical and analog approaches in general terms, we focus on identifying a concrete problem relevant for quantum materials and a parameter regime where the QPU may achieve advantages over state-of-the-art classical methods, whether in run-time performance or energety efficiency.

For the classical baseline in our comparison, we consider matrix product state (MPS) based methods, as they represent the most widely used and well-developed numerical approach for simulating quantum systems of medium size, systems consisting of tens of qubits~\cite{hauschild2018efficient,fishman2022itensor, emu-mps}. MPS techniques have a long track record of accuracy and efficiency for medium-sized systems, and their performance characteristics are well understood, making them a suitable benchmark for assessing quantum hardware. As an example of an emergent classical method, we also include neural quantum states (NQS)~\cite{carleo_solving_2017}, which have recently demonstrated comparable performance relative to established tensor-network approaches in certain problem settings~\cite{schmitt2025simulatingdynamicscorrelatedmatter}. Unlike MPS, the practical limits of NQS methods remain less clearly defined, offering an interesting point of comparison as their capabilities continue to be explored.

The remainder of this paper is structured as follows. Section~\ref{Sec_PC} defines the computational problem addressed in this work. Sections~\ref{sec:Qresources},~\ref{sec:Cresources} and ~\ref{sec:ScalingC} compare computation times between employing a neutral atom analog QPU and classical simulation methods for simulating post-quench dynamics. Sections~\ref{sec:Energy_comparison} and ~\ref{sec:extrap} present an analysis of energy consumption across these methods. Finally, Section~\ref{sec:outlook} summarizes our findings and outlines directions for future research.

\section{Problem choice}\label{Sec_PC}

We consider an Ising-like Hamiltonian on the square lattice, a problem of relevance for many applications in quantum materials~\cite{sachdev2023quantum}. This problem has been extensively studied both theoretically and experimentally for its ground state properties in the antiferromagnetic regime~\cite{Scholl_2021, Semeghini_2021,ebadi2021quantum,zhang2025probing}, and for its dynamics that can lead to exotic phenomena such as quantum scars~\cite{bluvstein2021controlling}, coarsening dynamics~\cite{manovitz2025quantum}, confinement and roughening dynamics~\cite{pavevsic2025constrained,krinitsin2025roughening} and Kibble-Zurek mechanisms~\cite{schmitt_quantum_2022}. In particular, Refs.~\cite{choi2023preparing,shaw2024benchmarking} explored post-quench dynamics regimes where the distribution of the quantum state in the Z-basis follow a Haar random distribution and showed an advantage in accuracy of sampling such states on a neutral atom QPU with respect to their classical counterpart. While the ground state properties of this stoquastic Hamiltonian can be well captured by classical methods, its dynamics can be very complex for parameter regimes that are non perturbative~\cite{vovrosh2025simulatingdynamicstwodimensionaltransversefield}. We therefore consider a quench compatible with our Rydberg atom quantum processing unit.

In a typical setup $^{87}$Rb atoms are first loaded into a magneto-optical trap, then transferred to individual optical tweezers generated using spatial light modulators or acousto-optic deflectors and typically separated by a few microns. In the analog mode, the qubit states are encoded as $\ket{g} = \ket{5S_{1/2}, F=2, m_F=2}\equiv \ket{0}$ and $\ket{r} = \ket{60S_{1/2}, m_J = 1/2}\equiv \ket{1}$.These two states form an effective two-level system, allowing us to map the neutral atom qubits onto spin-$1/2$ particles. 
The resulting system is governed by 
\begin{equation}\label{eq:ryd}
H = \sum_{i<j} V_{ij} n_i n_j + \frac{\Omega}{2} \sum_i \sigma_i^x - \Delta \sum_i n_i,
\end{equation}
where $\sigma_i^{x/z}$ are the Pauli-$x/z$ operators acting on site $i$ and $n_i = \frac{1}{2}(1+\sigma_i^z)$ is the projector onto the Rydberg state of qubit $i$. Here, $\Omega$ is the Rabi frequency of the laser driving the $\ket{0}\leftrightarrow \ket{1}$ transition, and $\Delta$ is the detuning of the laser from the Rydberg transition. The interaction term $V_{ij}$ describes the van der Waals interaction between atoms in the Rydberg state, typically scaling as $V_{ij} = C_6/|r_i - r_j|^6$, where $C_6$ is the interaction coefficient and $|r_i - r_j|$ is the distance between atoms $i$ and $j$.

This Hamiltonian captures the essential physics of Rydberg-mediated interactions in neutral atom arrays, providing a versatile platform for exploring strongly correlated quantum dynamics. Throughout this work, we consider the time evolution (post-quench) of the initial $\ket{00\cdots 0}$ state under the time-independent Hamiltonian with parameters $\Omega/2\pi = 2$MHz, $\Delta = \frac{1}{2}\sum_jV_{0j}$ in which $0$ is the index of a central atom in the square lattice with nearest neighbor atomic separation $R=\left(C_6h_x/\left(2\Omega\right)\right)^{1/6}$ such that $h_x=2.5$. These are chosen in accordance with Ref.~\cite{vovrosh2025simulatingdynamicstwodimensionaltransversefield}

\begin{figure*}
    \centering
    \includegraphics{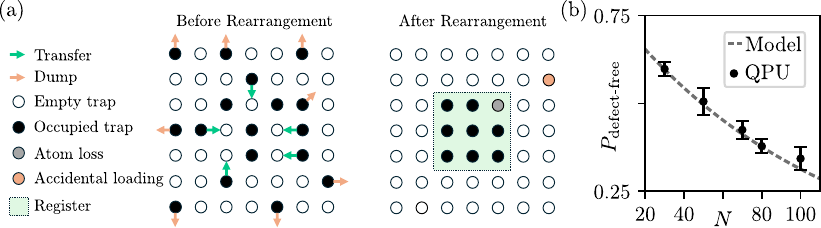}
    \caption{(a) Illustration of the different processes required during the rearrangement step. (b) Comparison between the mathematical model (dashed line) describing the probability to have zero defects in the register and experimental data (dots), $N$ being the number of atoms in the register. Here, the error bars are given by the standard deviation of the defect-free distribution over time.}
    \label{fig:Fig_2}
\end{figure*}

It is important to note that QPUs are noisy and this limits their coherence time. While the precise characterization of the noise is beyond the scope of this paper, we would like to emphasize key points. Neutral atoms are noisy but they are very well isolated systems allowing for a deep understanding of dominant noise sources. By controlling the noise, it has been demonstrated that these QPUs can reach regimes where classical methods are not accurate anymore for global quantities such as the fidelity~\cite{shaw2024benchmarking}. In this work, we are considering local observables and in that case the errors due to the noise remain bounded with the increase of the system size~\cite{daley2022practical,cai2023stochastic,schiffer2024quantum}. We therefore expect that the combination of experimental control of the noise, the increase in the system size and some potential error mitigation can allow the measure of local observables with bounded errors.

\section{Resource estimation for quantum simulation}\label{sec:Qresources}

In this section, we estimate the total time required to simulate the quantum dynamics described in Section~\ref{Sec_PC} using a neutral atom QPU such as Orion Alpha. Extracting meaningful physical information, such as the expectation value of a local observable, typically requires many shots to build sufficient statistical accuracy.. The key complication is that not all shots executed on current quantum hardware yield usable data. In practice, imperfections during initialization or rearrangement can lead to defects, whose mechanisms are sketched in Fig.~\ref{fig:Fig_2}(a).  

To address this issue, we model an effective zero-defect rate and analytically derive the scaling behavior of the total QPU simulation time, incorporating contributions from both the required number of defect-free shots and imperfections in the hardware.
This allows for a more realistic estimate of the time cost associated with executing the full protocol on a quantum device. 

We begin by estimating the probability that a given shot has no defects. The initial loading of atoms into optical tweezers is intrinsically probabilistic. Because of light-assisted collisions, if two atoms are captured in the same trap, both are lost, leading to a maximum filling probability of $50\%$ per site. As a consequence, the initial array of traps (the layout) should be larger than the register we aim to prepare (at least two times larger): it contains numerous vacancies, and a rearrangement step is required in order to prepare the desired configuration (the register) in which every site is occupied by a single atom~\cite{Barredo_2016, Schymik_2020}. This rearrangement process consists in moving atoms from filled sites to empty ones until the register is fully populated, and dumping any additional atoms. The success of this procedure can be quantified by the probability of obtaining a defect-free register, i.e., an array with no missing atoms after rearrangement and removal of unwanted atoms. Several processes contribute to this probability; we now describe them in turn.

The first contribution arises from the transfer of the atoms: each atom that needs to be relocated has a probability $p_{\text{transf}}$ of being successfully picked up, transported, and released at its target site. For $\text{N}_{\text{transf}}$ such operations, the contribution is $(p_{\text{transf}})^{\text{N}_{\text{transf}}}$.

The second contribution comes from the imperfect dumping of atoms lying outside the register. Removing such atoms requires successfully picking them up, which occurs with probability $p_{\text{pick-up}}$. For $\text{N}_{\text{dump}}$ dumps, the success factor is $(p_{\text{pick-up}})^{\text{N}_{\text{dump}}}$.

In addition, traps not involved in either transfers or dumps may still suffer from accidental loading of atoms with a probablity is $p_{\text{acci}}$. Hence, the probability that no accidental events occur in the $\text{N}_{\text{traps}} - \text{N}_{\text{transf}} - \text{N}_{\text{dump}}$ idle traps is $(1 - p_{\text{acci}})^{\text{N}_{\text{traps}} - \text{N}_{\text{transf}} - \text{N}_{\text{dump}}}$.

Finally, atoms within the register may be lost during this process due to collisions with background gas or imperfect detection with probability $p_{\text{loss}}$. For the $\text{N}_{\text{register}} - \text{N}_{\text{transf}}$ atoms that were not moved, the survival probability is $(1 - p_{\text{loss}})^{\text{N}_{\text{register}} - \text{N}_{\text{transf}}}$.

Assuming these processes are statistically independent, the total probability of preparing a defect-free register is then given by the product of these contributions
\begin{equation} \label{zero_defect}
\begin{aligned}
&P_{\text{defect-free}} =\;  (p_{\text{transf}})^{\text{N}_{\text{transf}}}
(p_{\text{pick-up}})^{\text{N}_{\text{dump}}} \times \\
& (1 - p_{\text{acci}})^{\text{N}_{\text{traps}} - \text{N}_{\text{transf}} - \text{N}_{\text{dump}}}
(1 - p_{\text{loss}})^{\text{N}_{\text{register}} - \text{N}_{\text{transf}}}.
\end{aligned}
\end{equation}
These quantities can be extracted experimentally by comparing pre- and post-rearrangement images while taking into account the moves computed by the rearrangement process predetermined by the Hungarian algorithm~\cite{Lee_2017}.

In order to verify the validity of the phenomenological model introduced above, we perform measurements on our QPU using an layout consisting of $200$ traps and varying the register size between $30$ and $100$ atoms. For each configuration, we extract the relevant observables by comparing the pre- and post-rearrangement images. Averaging over the different register sizes, we find~\footnote{These values are quoted without correcting for vacuum and detection losses.}
\begin{equation}
\begin{aligned}
& p_{\text{transf}} = (98.9 \pm 0.1)\%, \quad 
& p_{\text{pick-up}} = (99.8 \pm 0.1)\%, \\
& p_{\text{acci}} = (0.09 \pm 0.02)\%, \quad 
& p_{\text{loss}} = (0.9 \pm 0.1)\%.
\end{aligned}
\end{equation}
Figure~\ref{fig:Fig_2}(b) shows the comparison between the prediction of Eq.~\eqref{zero_defect} and direct measurements of the defect-free probability, obtained by counting the fraction of experimental realizations in which all register sites were occupied and all traps outside the register were successfully emptied. The excellent agreement between the experimental data and the analytical model demonstrates that Eq.~\eqref{zero_defect} captures the essential processes governing the preparation of defect-free registers in our system.

Given Eq.~\eqref{zero_defect}, we are now in a position to estimate the total simulation time required to accurately measure a local observable of a prepared quantum state. Specifically, consider a scenario in which we prepare a quantum state $\ket{\psi}$ and aim to estimate expectation values of the form $\langle \sigma^z_i \rangle$ and $\langle \sigma^z_i \sigma^z_j \rangle$, where $i$ and $j$ label specific qubits in the system. Since both $\sigma^z_i$ and $\sigma^z_i \sigma^z_j$ have eigenvalues $\pm 1$, the shot noise in these measurements can be modeled using a binomial distribution. Let $p$ denote the probability of measuring the outcome $+1$. Then, after $n$ measurement shots, the estimated expectation value of an observable $\hat{O} \in \{\sigma^z_i, \sigma^z_i \sigma^z_j\}$ is given by $\langle \hat{O} \rangle = 2p - 1$, with an associated standard deviation $\sigma_O = 2\sqrt{p(1 - p)/n}$.

To achieve a precision of $\alpha$ with 95\% confidence, we require that the standard deviation satisfy $\sigma_O \leq \alpha/2$. Solving for the number of required shots under this constraint yields
\begin{equation}
n = \frac{16p(1 - p)}{\alpha^2}.
\end{equation}
In the worst-case scenario, where $p = 0.5$, and for a tolerance $\alpha = 0.05$, we find that $n = 1600$ shots are needed. However, this estimate assumes that all shots are usable, which is generally not the case. As discussed above, imperfections in the filling of the atomic register can lead to defects, such as missing atoms, that render some shots unusable. We now extend our estimate to account for this zero-defect probability.

Assuming the probability of obtaining a defect-free register is given by Eq.~\eqref{zero_defect}, and aiming to collect $m$ usable (defect-free) shots, the probability of obtaining at least $m$ such shots out of a total of $n$ attempts follows a binomial distribution
\begin{equation}
\sum_{k=m}^n \binom{n}{k} P_{\text{defect-free}}^{k}(1 - P_{\text{defect-free}})^{n - k}.
\end{equation}
With this, given a system size $N$, we are able to calculate the number of shots required to accurately estimate a local observable and, with an assumed shot frequency of 1Hz, the corresponding run time. In the conclusions and outlook, we discuss potential technological improvements to reduce this run time.

\section{Resource estimation for classical simulations}\label{sec:Cresources}

\begin{figure}
    \centering
    \includegraphics{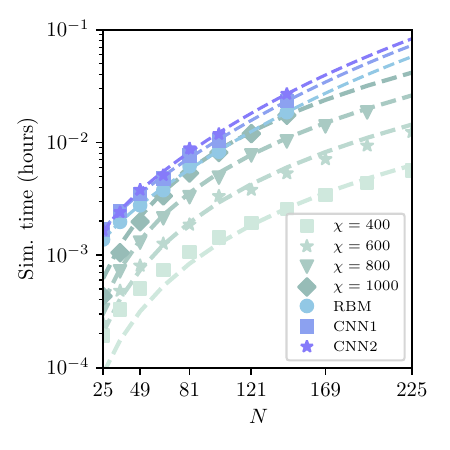}
    \caption{Run-time of a single 1 ns time step ($dtJ\sim0.01$, in which $J$ is the energy scale of the system) for a quantum quench to the Ising model, measured on an NVIDIA A100 GPU, as a function of system size. Results are shown for both MPS with varying bond dimensions in teal and NQS with different types of networks (containing varying numbers of parameters) in purple. Theoretical scaling fits are overlaid and show good agreement with the data, see App.~\ref{app:fitting} for details.}
    \label{fig:Fig_3}
\end{figure}

Several computational approaches have been developed to study the time evolution and ground state properties of many-body quantum systems. In this work we consider Matrix Product State (MPS) based methods as they are the most widely adopted ansatz for simulating quantum quenches in moderately-sized systems~\cite{Paeckel_2019}.  We also consider a time-dependent variational Monte Carlo (tVMC) approach based on neural networks (e.g. restricted Boltzmann machines or Convolutional Neural Networks) to represent many-body wave functions, commonly known as the Neural Quantum States (NQS) approach~\cite{carleo_solving_2017,schmitt2025simulatingdynamicscorrelatedmatter}.

We begin by deriving the dominant contributions of computational complexity and run-time scaling for both MPS and NQS simulations. This analytical groundwork provides insight into the expected computational cost as a function of key problem parameters. To validate and calibrate these theoretical expressions, we fit them to empirical data obtained from simulations performed on a single NVIDIA A100 40Gb SXM GPU for MPS due to the difficulty of parallelization (see App.~\ref{app:cpu} for a comparison of CPU and GPU run-times).
For NQS simulations, we perform simulations with multiple of these GPUs in certain cases (we comment more on this below). This approach enables us to generate accurate and practically relevant estimates of the performance of such simulations, which are valid in different regimes of post-quench dynamics.

\subsection{MPS simulations}
We analyze the key operations involved in the time evolution of a quantum state in MPS form to estimate the computational complexity. The total evolution is partitioned into $n$ discrete time steps, each executed using the two-site Time-Dependent Variational Principle (TDVP) algorithm \cite{Haegeman_tdvp_1, Haegeman_tdvp_2}, which enables an accurate treatment of the long-range interactions described by Eq.~\eqref{eq:ryd}.
Since each time step involves a similar sequence of operations, the total computational cost can be estimated by analyzing the cost associated with a single representative step and subsequently scaling by $n$. To quantify the per-step cost, we examine the dominant computational operations required to propagate the state.
Two contributions dominate the run-time: the construction of the effective environments~\cite{crosswhite_mpo, paeckel_mps} (also referred to as the left and right “baths”) and the application of the effective two-site Hamiltonian using the Lanczos algorithm~\cite{lancz_matrix_exp}.

The construction of the environment requires contracting the MPS tensors with the Matrix Product Operator (MPO) representation of the Hamiltonian in Eq.~\eqref{eq:ryd}. 
In a two-dimensional square lattice, each bulk qubit is most strongly coupled to its four nearest neighbors—two along the horizontal and two along the vertical directions—while additional weaker couplings to more distant sites arise from the long-range nature of the interactions. When mapped to a 1D MPO representation of the Hamiltonian, the two vertical couplings correspond to strong quasi–long-range interactions. 
As a result, for a square lattice of $N$ qubits, the MPO bond dimension scales with system size as $h = \mathcal{O}(\sqrt N)$, see App.~\ref{app:mps_details} for further details.

It follows that bath construction or update involves tensor contractions between the bath and an MPS tensor, contributing terms that scale as $\mathcal{O}(\sqrt N \chi^3)$, and contractions between the resulting intermediate tensor and an MPO tensor, which scale as $ \mathcal{O}(N \chi^2)$. Moreover, since the left and right baths must each be computed approximately $N$ times, the overall expected complexity is $\mathcal{O}(N^{3/2} \chi^3 + N^2 \chi^2)$

Following bath construction, the application of the effective two-site Hamiltonian involves performing a Krylov subspace evolution of the two-site tensor using the Lanczos algorithm. This process is typically iterated $k$ times per step and includes additional tensor contractions and a singular value decomposition to restore the canonical MPS form. These operations produce contributions with similar scaling, such that the total cost of one TDVP step can be modeled as the sum of both dominant terms. Aggregating all contributions and absorbing constant prefactors and repeated Krylov steps into fitted coefficients, the expected run-time per TDVP step can be expressed as
\begin{equation}
\langle \Delta t \rangle \approx \alpha N^{3/2} \chi^3 + \beta N^2 \chi^2,
\end{equation}
where $\alpha$ and $\beta$ are constants determined by hardware characteristics and algorithmic implementation details. The first term dominates in regimes where the entanglement, and hence the bond dimension $\chi$, is large, while the second term becomes significant in systems with modest entanglement but large system size $N$. This model provides a predictive framework for assessing the computational demands of MPS simulations and for informing optimization strategies, such as bond dimension truncation or interaction locality adjustments, to improve run-time efficiency.

\subsection{NQS simulations}

We now discuss the expected system-size scaling of the complexity (and the corresponding run-time) of the NQS approach.  More details about the method are provided in the Appendix and in Refs.~\cite{schmitt2025simulatingdynamicscorrelatedmatter,Schmitt2022,Vicentini2022}. Here, instead, we summarize the key aspects of the tVMC algorithm that determine its computational complexity.

Similarly to tensor network approaches, the NQS aims to represent the amplitudes of quantum states, parametrized by a set of variational parameters, $\vec{\theta} = (\theta_1,\theta_2,...,\theta_{N_v})$, 
taking the form of a neural network, with $N_v$ denoting the total number of variational parameters. The dynamics of the system is described by updating the set $\{ \theta_k \}$ at each time step by solving the first-order differential equation
$S_{k,k^{\prime}} \, \dot{\theta}_{k^{\prime}} = -i F_k$,
where the matrix $S$ (called the Fisher matrix) has dimension $N_v \times N_v$, and $F$ (the force vector) has dimension $N_v$.  

The complexity of NQS simulations is thus related to  the computational cost of evaluation  (i) $S$ and $F$, and (ii) the required inversion of the $S$ matrix to solve the differential equation.  

On the one hand,  the gradients and the  local energy defining $S$ and $F$  is evaluated by Monte Carlo sampling. For this, one must generate $N_{\mathrm{MC}}$  samples with, as for example (and considered) by using the Metropolis Algorithm.
For each MC sample, a forward-pass evaluation of the neural network is required, whose complexity ${\cal C}_s$ is characterized by the number of floating-point operations (FLOPs). To leading order, the total complexity related to this part is then given by  
${\cal O}(N_{\mathrm{MC}} N) \times {\cal C}_s$, where we assume that for each MC sample one typically consider $N$ local updates.
Other aspects can also affect the complexity. For instance, when the Hamiltonian contains a number of non-diagonal terms that scale as ${\cal O}(N^\alpha)$ (with $\alpha > 1$), additional overhead is introduced to access the local energy.  

On the other hand, the complexity of inverting the matrix $S$ scales as  
${\cal O}(N_v^3)$.

In the simulations considered here, the complexity of part (i) is typically the dominant contribution, since we work in the regime $N_{\mathrm{MC}} \gg N_v$. 
To leading order, the number of FLOPs required to perform a single time step with the tVMC approach thus scales as ${\cal O}(N_{\mathrm{MC}} N) \times {\cal C}_s$. 
Assuming that the number of variational parameters scales as $N_v = \alpha {\cal O}(N)$, where $\alpha$ is a parameter that depends on the particular architecture of the NQS,  and taking into account that for shallow convolutional neural network (CNN) or restricted Boltzman machine (RBM) ${\cal C}_s = {\cal O}(N^2)$, we obtain that the complexity of the NQS algorithm follows the scaling: ${\cal O}(\alpha N_{\mathrm{MC}} N^3)$. 

It is worth mentioning that the actual run time of the simulations depends on additional factors that are not directly captured by the complexity analysis above. Examples include the batch size of samples used during the neural network forward pass and the number of GPUs available to perform independent Monte Carlo sampling in parallel; in the latter case, one often observes an approximately linear speedup with the number of GPUs~\cite{Schmitt2022}. As we discuss in the following section, empirical results for the run time of tVMC simulations with shallow-CNN or RBM architectures are well described by a polynomial function of 
$N$ of degree three.

\subsection{Results}

In Fig.~\ref{fig:Fig_3}, we present simulation results of the quench protocol described in Section~\ref{Sec_PC}, exploring how the computational complexity scales for both MPS and NQS methods. Specifically, we report the average run time required to compute a single time step, with a resolution of approximately $dt=1$ns, for each method across varying system parameters. Note, this is such that $dtJ \sim 0.01$ where $J=C_4/4r^6$ is the energy scale of the system, in which $r$ is the nearest neighbor separation in the square lattice. Superimposed on this data are the corresponding theoretical fits derived previously, which show remarkably good agreement with the observed run times, lending strong support to the validity of our scaling analysis and allowing for predictions beyond what is possible with currently available hardware.

However, before we can directly compare classical and QPU simulation times, it is important to note that classical methods do not always yield exact results, their accuracy depending on the complexity of the quantum state being simulated. Therefore, in the following section, we introduce convergence criteria to determine the minimum classical resources required to achieve reliable results.

\section{Scaling of converged classical simulations}\label{sec:ScalingC}

To assess the validity of both MPS and NQS simulations, we employ independent convergence criteria tailored to each method. For MPS, we rely on physically motivated indicators: convergence is assumed when the total energy of the system is conserved and the underlying symmetries of the lattice, such as translational invariance or spin conservation, are preserved throughout the simulation. These criteria ensure that the MPS faithfully captures the essential features of the quantum state. 
For NQS, we consider estimations of the TDVP error.

In this section, we shift our focus from the simulation time required for a single small time step to the total time needed to simulate a quench protocol over a given pulse duration. That is, rather than considering the cost of advancing the wavefunction by $dt \approx 1$ns, we evaluate the cumulative computational resources required to prepare the full wavefunction and measure an observable of interest at varying times $t$ after the quench. This approach allows us to capture the scaling of the total simulation effort as a function of the evolution time, providing a more direct comparison with the duration of quantum experiments.

\subsection{Benchmarking MPS against QPU run time}

For the problem considered in Section~\ref{Sec_PC}, we can impose physically motivated criteria to identify whether classical simulations yield correct results. Since we are simulating a closed quantum system undergoing a quench to a time-independent Hamiltonian, energy should be conserved throughout the evolution. Additionally, the Hamiltonian is spatially uniform, which implies that the dynamics must respect the symmetries of the underlying square lattice, i.e. the D\textsubscript{8} symmetry group. These constraints provide concrete benchmarks against which to test convergence.

To quantify symmetry preservation, we define the symmetry error of an MPS simulation as the maximum difference between the observable matrix and any of its symmetry-equivalent counterparts under the D\textsubscript{8} dihedral group. That is, \begin{equation}
\epsilon_{\text{D}_8} = \max_{g \in D_8} \left\| \langle \hat{O} \rangle - g \cdot \langle \hat{O} \rangle \right\|,
\end{equation}
where $g$ denotes an element of the D\textsubscript{8} symmetry group acting on the observable matrix $\langle\hat{O}\rangle$. We then impose two convergence conditions on MPS simulations: (1) the relative deviation from energy conservation must remain below 5\%, and (2) the D\textsubscript{8} symmetry error must be less than 40\%. Given any results that obey these criteria, we consider the results accurate, see Ref.~\cite{vovrosh2025simulatingdynamicstwodimensionaltransversefield} for details. 

\begin{figure}
    \centering
    \includegraphics{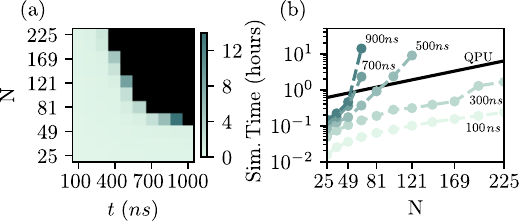}
    \caption{(a) Minimum simulation time required for MPS to achieve convergence as a function of pulse duration $t$ and system size $N$. The black region highlights the pulse durations in which 40Gb of RAM was not enough to achieve convergence. (b) Comparison of total simulation time between classical MPS and estimates for QPU execution across varying system sizes for increasing pulse durations.}
    \label{fig:Fig_4}
\end{figure}

In Fig.~\ref{fig:Fig_4}(a), we present the minimum required simulation times for MPS simulations of the quench protocol under consideration, as a function of pulse duration. For each pulse duration, we perform a series of simulations with progressively increasing bond dimensions, in increments of 100, until convergence was achieved according to the criteria described above. The reported simulation time corresponds to the minimum bond dimension that satisfies the convergence requirements, thereby representing the lower bound on classical simulation time for that parameter regime. The black region indicates where simulations could not be converged due to memory limitations on a 40Gb RAM A100 GPU, highlighting the growing computational cost in more entangled regimes.

The results in Fig.~\ref{fig:Fig_4}(a) enable a direct comparison between MPS and QPU simulation times across varying system sizes, with the corresponding data summarized in Fig.~\ref{fig:Fig_4}(b). For short post-quench dynamics, in which the total entanglement in the system remains low and tensor network methods are applicable, we observe that classical MPS simulations consistently outperform the QPU, even for the largest system sizes considered, and are expected to continue doing so beyond this range due to their more favorable scaling behavior. However, for medium-duration post-quench dynamics ($\ge500$ns), the performance of MPS deteriorates rapidly with system size, while the QPU maintains a relatively constant execution time, leading to a clear advantage in favor of the QPU. Considering that the coherence time of a neutral atom QPU operating in analog mode can reach up to $6\mu$s, these results highlight a substantial regime in which quantum hardware can outperform classical MPS methods.

\subsection{Comparison between NQS and QPU run time}

Time-dependent VMC  simulations provide a promising alternative to tensor network approaches such as MPS, given their ability to efficiently represent certain highly entangled states and 
the parallelizable implementation of tVMC codes that typically allows for an almost linear reduction in run-time with the number of GPUs \cite{Schmitt2022}.  
It remains unclear, however, whether tVMC simulations can reach convergence for the particular type of quench considered in this work and time scales of order $\approx 4000$\,ns expected for neutral-atom QPUs. Recent studies suggest that the observed limitations in capturing the dynamics in post-quench dynamics are not due to an intrinsic lack of representability of NQS, but rather arise from numerical instabilities in solving the TDVP \cite{schmitt2025simulatingdynamicscorrelatedmatter} equation or from finite Monte Carlo sampling \cite{Sinibaldi_2023}.  

To assess the convergence of NQS simulations we consider the following quantity
\begin{equation}
r^2(t) = \frac{{\cal{D}}(\ket{\psi_{\vec{\theta}(t + \delta t)}}, e^{-i\hat H \delta t}\ket{\psi_{\vec{\theta}(t)}})^2}{{\cal{D}}(\ket{\psi_{\vec{\theta}(t + \delta t)}},\ket{\psi_{\vec{\theta}(t)}})^2},
\label{tdvp_err}
\end{equation}
where ${\cal{D}}$ is the Fubini-Study distance between two quantum states \cite{carleo_solving_2017,schmitt2025simulatingdynamicscorrelatedmatter}. The numerator of the expression mentioned above represents the optimization objective used to define the equation of motion for the variational parameters $\vec{\theta}$, and the denominator is introduced as a natural scale.
The $r^2$ can be obtained with Monte Carlo sampling by considering a second other approximation \cite{schmitt2025simulatingdynamicscorrelatedmatter}.
 The integrated error up to a time $t$, i.e. $R^2(t) = \int_0^t r^2(t^{\prime})dt^{\prime}$, can be used as a figure of merit for the convergence of the NQS results. 
 
For a given system size $N$, the simulations are performed with a fixed NQS architecture. 
As a reference of NQS architecture, we consider a two-layer CNN containing a number of variational parameters that scale as $O(N)$ \cite{schmitt2025simulatingdynamicscorrelatedmatter}. 
In the SM (see Fig.~\ref{fig1_sm}) we present the colormap of $R^2$ as a function of system size and evolution time. 
The results indicate that tVMC simulations with CNN employed face difficulties in reaching convergence for intermediate post-quench durations, specifically for $t > 400$ ns (i.e. $tJ \approx 1.0$). 
This behavior is reflected in the rapid growth of $R^2$ with time, see App.~\ref{app1}. 

In panel (a) of Fig.~\ref{fig:Fig_5}, we highlight only the regime of parameters that can be considered reliable (i.e., cases for which $R^2 < 0.05$). 
A more precise discussion of the convergence of NQS results for different observables and architectures is presented in Ref.~\cite{vovrosh2025simulatingdynamicstwodimensionaltransversefield}. 

Panel (b) of Fig.~\ref{fig:Fig_5} reports a direct comparison between NQS simulations and QPU run-time estimates, with NQS simulations executed on $4$ NVIDIA A100 40GB SXM GPUs. 
Although NQS approaches preserve polynomial run-time scaling, analogous to that of the QPU, our results indicate that, already at intermediate post-quench durations, the QPU exhibits a clear run-time advantage for the estimation of local observables.

\begin{figure}
    \centering
    \includegraphics{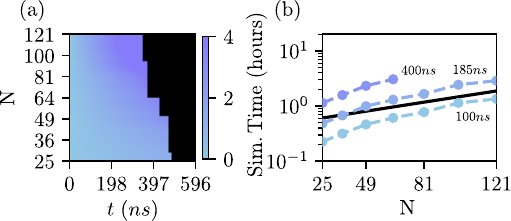}
    \caption{(a) Minimum simulation time required for NQS to achieve convergence as a function of quench durantion and system size. The black region highlights the quench durantions in which simulations do not converge. (b) Comparison of total simulation time between classical NQS and estimates for QPU execution across varying system sizes for increasing pulse durations. Only simulations with $R^2$ below a certain threshold (see text) are considered.}
    \label{fig:Fig_5}
\end{figure}

\section{Energy consumption}
\label{sec:Energy_comparison}

In this section, we analyze the power requirements of both MPS and NQS classical simulations and quantum simulation executed on a neutral atom QPU in analog mode. By comparing their respective energy footprints, we aim to assess which approach is more efficient as system size and simulation complexity scale up.

While simulation time is a key performance metric, it is not the only factor to consider when evaluating different simulation approaches. Power consumption is also a critical consideration, particularly in the context of large-scale simulations where energy efficiency can significantly impact both operational cost and sustainability. 

For both the MPS and NQS methods, we measure the average power consumption of an A100 GPU during simulation runs across a range of ansatz complexities, Fig.~\ref{fig:Fig_6}(a-b). The estimations is performed by monitoring power consumption with the NVIDIA command \textit{nvidia-smi dmon}, which logs GPU utilization, memory usage, and power draw throughout the simulation. As expected, the average power usage increases with the numerical complexity of the task, reflecting the higher computational demands of more entangled or less compressible quantum states. Importantly, this measurement captures only the power drawn by the GPU as reported by \textit{nvidia-smi} monitoring tool and does not include the energy consumed by the host CPU system, memory, cooling infrastructure, or network interconnects required to run the simulation. 

To estimate the power consumption of the QPU, we use the data presented in Fig.~\ref{fig:Fig_6}(c), which is based on experimental measurements collected by all power distribution units within the QPU system over a two-month period; The average of these measurements is taken as the representative power draw of the device during typical operation. This measurement includes the power consumption of all optical cores, electronic racks, laser systems, and laser control racks. Consequently, our comparison is slightly biased in favor of the GPU-based simulations, as it does not account for their full system-level power footprint, whereas the QPU estimate reflects the complete device power draw during typical operation. Nevertheless, this distinction will not affect the main conclusion of the following section: due to the QPU’s favorable run-time scaling, it still consumes less total energy than classical simulation methods for the same post-quench dynamics task for medium duration times.

\begin{figure}
    \centering
    \includegraphics[width=1.0\linewidth]{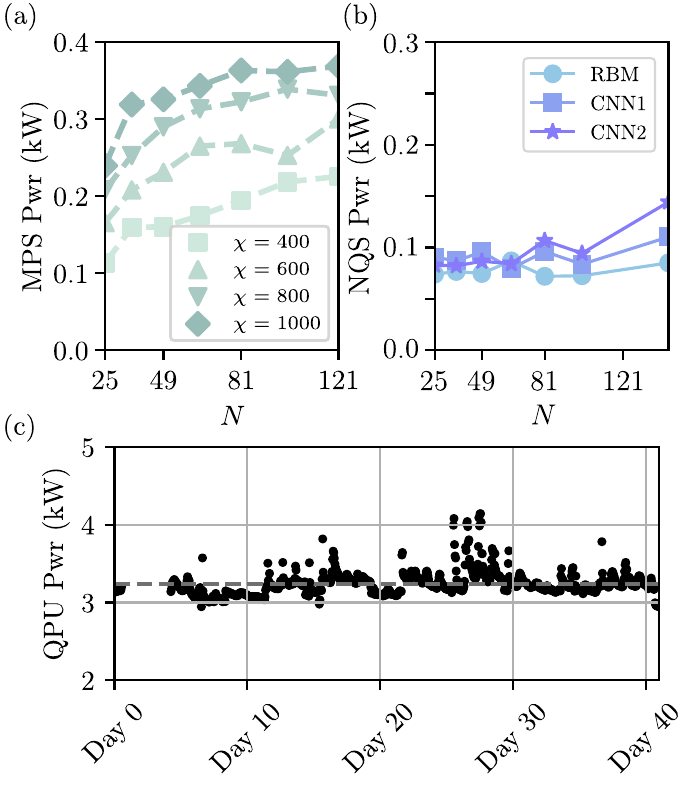}
    \caption{(a-b) Average power consumption of the NVIDIA A100 GPU during MPS and NQS simulations with increasing system size for various bond dimensions. (c) Total power consumption of a neutral atom QPU operating in analog mode.}
    \label{fig:Fig_6}
\end{figure}

\begin{table*}[t]
\centering
\renewcommand{\arraystretch}{1.15}
\begin{tabular}{l *{9}{c}}
\toprule
Problem size 
& \multicolumn{3}{c}{15$\times$15} & \multicolumn{3}{c}{20$\times$20} & \multicolumn{3}{c}{25$\times$25} \\
\cmidrule(lr){2-4}\cmidrule(lr){5-7}\cmidrule(lr){8-10}
& Mem & Time & Energy (kWh)
& Mem & Time & Energy (kWh)
& Mem & Time & Energy (kWh) \\
\midrule
Classical MPS ($\chi=1000$) & 150 GB & 30 d & 0.29k & 600 GB  & 140 d  & 1.4k   & 1.4 TB  & 1.0 y  & 3.6k \\
Classical MPS ($\chi=2000$) & 360 GB & 93 d & 0.88k    & 1.4 TB  & 1.1 y  & 4.0k  & 3.2 TB  & 2.9 y & 10k \\
Classical MPS ($\chi=3000$) & 700 GB  & 220 d  & 2.1k   & 2.8 TB  & 2.7 y & 9.3k  & 6.3 TB & 6.6 y & 23k \\
\textbf{QPU}         & --   & \textbf{6.3 h}  & \textbf{20}    & --   & \textbf{48.3 h} & \textbf{156}  & --   & \textbf{27.5 d} & \textbf{2k} \\
\bottomrule
\end{tabular}
\caption{Resource estimation for simulations of large lattice sizes with both MPS methods and a neutral atom qpu in analog mode for apulse duration of $4\mu$s, or $\sim10J$. (Mem = memory, Energy = energy). Units: y = years, d = days, h = hours.}
\vspace{-10pt}
\label{tab:resources-condensed}
\end{table*}

\section{System-size extrapolation of run time and energy consumption}\label{sec:extrap}

In this section, we discuss the extrapolation of resource requirements for classical simulations and contrast them with QPU estimates. 

This analysis is possible for MPS simulations since their errors can be systematically decreased to an arbitrarily low level by enlarging the bond dimension, albeit at the expense of higher computational cost and memory requirements. 
Given that the same level of control is not currently understood for the NQS approach, in the following we focus exclusively on extrapolations of MPS simulations.

By combining the fitted run-time data from Fig.~\ref{fig:Fig_4} with the known memory scaling behavior of MPS methods (see the supplementary material), we can extrapolate both simulation times and memory requirements for large-scale simulations. Additionally, assuming that the GPU operates at its maximum rated power consumption of 0.4 kW under heavy computational load~\cite{nvidia_a100_datasheet}, we can estimate the total energy usage for these simulations. This assumption is supported by the observed increase in GPU power consumption with larger $N$ and bond dimension, as illustrated in Fig.~\ref{fig:Fig_6}(a). In Table.~\ref{tab:resources-condensed}, we present projected simulation times, memory demands, and power consumption for a 4$\mu$s pulse duration ($\sim10J$) applied to large square lattices, evaluated across a range of bond dimensions. Even for moderately sized systems of $O\left(100\right)$ atoms and post-quench durations on the order of a few microseconds, our results clearly show that the QPU outperforms classical methods by several orders of magnitude in both run time and energy efficiency.

It is known that the entanglement generated after the quench grows approximately linearly with time across the system, leading to an exponential increase in the MPS bond dimension needed to accurately represent the state. In 2D lattices, this growth is further amplified by the width of the lattice, so that longer pulse durations or larger system sizes can require substantially higher bond dimensions than those considered here. Consequently, the projected simulation times, memory requirements, and energy consumption presented in Table~\ref{tab:resources-condensed} likely underestimate the true computational cost for very long pulse durations with MPS methods.

\section{\label{sec:outlook} Conclusions and Outlook} 
In this work, we have numerically demonstrated that the run-time scaling of both MPS and NQS methods for simulating post-quench dynamics of the long-range Ising model on square lattices is consistent with theoretical predictions. Leveraging experimental data from a neutral atom QPU in analog mode, we estimated the time required to perform equivalent simulations on quantum hardware.

Our analysis revealed that MPS and NQS classical simulations exhibit unfavorable scaling in both simulation time and energy consumption when convergence is strictly enforced. In contrast, the neutral atom QPU maintains favorable performance as system size and pulse duration increase for protocols in which devise noise remains bounded. Beyond such protocols, error-erasure techniques, which convert dominant physical errors into detectable erasures, will likely allow the QPU to maintain favorable scaling relative to classical methods even in the presence of higher noise levels~\cite{Wu_2022, holland2024demonstrationerasureconversionmolecular}.

By combining performance data from both classical GPU-based simulations and QPU experiments, we projected the memory requirements, run times, and energy costs associated with simulating large-scale quantum registers. These projections indicate that analog quantum devices offer a competitive, and in many regimes, superior, alternative to classical simulation methods. Notably, they can achieve comparable accuracy with significantly reduced computational overhead and power consumption, making them a promising platform for future large-scale quantum simulation tasks~\cite{gibson2025quantumcountingrydbergblockade}. It remains an important direction for future work to benchmark neutral atom QPUs against other classical methods such as Pauli Propagation~\cite{
broers2024exclusiveorencodedalgebraicstructure, broers2025scalablesimulationquantummanybody, PRXQuantum.6.020302, rudolph2025paulipropagationcomputationalframework}, Tree Tensor Networks~\cite{krinitsin2025timeevolutionquantumising, Pave_i__2025, Krinitsin_2025, pavešić2025scatteringinducedfalsevacuum} and two-dimensional Tensor Network approaches~\cite{PRXQuantum.5.010308, tindall2025dynamicsdisorderedquantumsystems, rudolph2025simulatingsamplingquantumcircuits, Begu_i__2024, 7jzt-xhn6} (2DTN). In particular, Ref.~\cite{vovrosh2025simulatingdynamicstwodimensionaltransversefield} shows that 2DTN methods can reliably capture the post-quench time dynamics of the nearest-neighbor transverse-field Ising model, achieving timescales comparable to those reached by the methods used in this work for the largest system sizes considered. Moreover, these methods are expected to scale more favorably with system size.

While our estimates provide a meaningful benchmark for the current capabilities of neutral atom QPUs, it is important to emphasize that they likely represent an upper bound on the true resource requirements. In practice, several experimental advances could further reduce overhead and improve efficiency, including the use of multiple rearrangement stages to enhance loading efficiency, averaging over symmetrically equivalent qubits to suppress statistical fluctuations, and the development of more sophisticated rearrangement algorithms to minimize atom loss and idle time~\cite{PhysRevA.108.023107, PhysRevA.108.023108, 2rds-k4n9, lin2024aienabledrapidassemblythousands}. Another key advance is the implementation of continuous atom reloading, which allows rapid recovery of lost atoms between experimental cycles without disrupting the rest of the register~\cite{li2025fastcontinuouscoherentatom,chiu2025continuousoperationcoherent3000qubit}. By maintaining high filling fractions with minimal dead time, continuous reloading can substantially increase the achievable shot rate, thereby reducing the overall wall-clock time required for post-quench dynamics experiments. Moreover, if one considers simulations that are intrinsically robust to a small fraction of defects, the scaling of QPU run-time can be significantly improved. In particular, for robustness levels on the order of $O(1\%)$ of the register, the run-time overhead can even become constant with system size. As these and other improvements are implemented, the performance gap between analog quantum devices and classical simulations is expected to widen, reinforcing the long-term promise of neutral atom platforms for scalable quantum simulation.

\acknowledgements
We thank Christophe Jurczak, Antoine Browaeys, Thierry Lahaye, Olivier Ezratty, Alexia Auff\`eves, and Victor Drouin Touchette for carefully reading the manuscript and providing insightful suggestions. We acknowledge funding from the European Union the projects PASQuanS2.1 (HORIZON-CL4-2022-QUANTUM02-SGA, Grant Agreement 101113690).

\appendix

\section{Further details of MPS simulations}
\label{app:mps_details}

In this section, we estimate the memory requirements for storing and evolving quantum states using MPS tensor networks. The primary contributors to memory consumption are: (i) the MPS tensors themselves, (ii) the left and right bath tensors, and (iii) the intermediate contraction tensors generated during time evolution. All subsequent analyses are performed using upper-bound estimates.

The first source of memory consumption is the MPS itself. An MPS consists of $N$ nodes, each represented by a three-dimensional tensor of size $(\chi, d, \chi)$, storing complex double-precision numbers of size $s$. Consequently, the total memory required to store the MPS is  
\begin{equation}
M_{\text{MPS}} = s d\chi^2 N.
\end{equation}
In our work, we define the MPS using a snaking path: starting from the bottom-left corner of the square lattice, we traverse rightward along the first row, then continue from the left of the second row, and so on.

Another significant source of memory consumption is the set of bath tensors.  
The left and right environment baths are obtained by contracting subsets of the MPS nodes $\psi_i$ with the corresponding tensors $H_i$ of the Hamiltonian $H$ in MPO form.  
Specifically, the left bath is defined as
\begin{equation}
L_i = \prod_{m=1}^{i} \psi_m^\dagger H_m \psi_m ,
\end{equation}
contracting tensors from the left, starting at site $m=1$ and proceeding to site $m=i$. The right bath is defined as
\begin{equation}
R_i = \prod_{m=i}^{N} \psi_m^\dagger H_m \psi_m.
\end{equation}
Tensor contractions can be performed starting from the rightmost tensor $m=N$ and proceeding toward site $m=i$, inclusive. 
Each bath tensor is a three-dimensional array of size $(\chi, h_i, \chi)$, 
where $h_i$ denotes the bond dimension between sites $(i, i+1)$ of the Hamiltonian in its MPO form, 
and $\chi$ is the bond dimension of the MPS. 
The value of $h_i$ varies between sites, as determined by the bond dimension profile of the Hamiltonian. 
The bond-dimension profile $h_i$ for square lattices, shown in Fig.~\ref{fig:bond_dim_prof}, 
typically exhibits a linear growth followed by a saturation region, 
reaching its maximum near the center of the system, with  $h_{\mathrm{max}} \approx 3 \sqrt{N} + 2$ for long-range interacting Hamiltonians, as illustrated in the inset of Fig.~\ref{fig:bond_dim_prof}.
\begin{figure}
    \centering
    \includegraphics{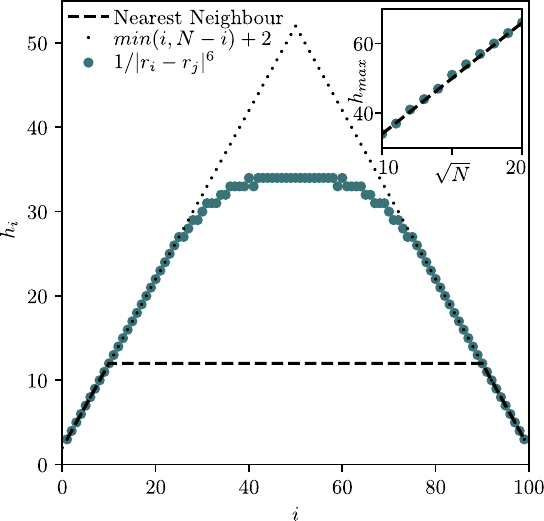}
    \caption{Bond-dimension profile of a Rydberg Hamiltonian in MPO form for a square lattice with $N = 100$, i.e. a $10\times10$ grid. The inset shows how the maximal bond dimension such an MPO depends on $N$.}
    \label{fig:bond_dim_prof}
\end{figure}

The total memory consumption for the baths is estimated as the sum of the contributions from individual bath tensors $\sum_i^N M(\chi h_i \chi) = \chi^2 \sum_i^N M(h_i)$. This sum is nothing but an area under the $h_i$ profile, that we approximate with trapezoid. Thus $\sum_i^N M(\chi, h_i, \chi) \approx \chi^2  M((3\sqrt{N}+2)(N - 3\sqrt{N}-2))$. The memory required to store baths reads
\begin{equation}
M_{\text{baths}} = s\chi^2 (3 N ^{3/2} - 7 N - 12N^{1/2}-4) \approx 3s\chi^2N^{3/2}
\end{equation}

A further source of memory consumption arises during time evolution of the MPS tensors 
using the two-site TDVP algorithm with the Lanczos method. In this approach, a Krylov subspace is constructed; forming a basis of $k$ vectors requires storing $k$ three-dimensional tensors of size $(\chi, d, \chi)$. In practice, the basis size is typically limited to $k_{\mathrm{max}} = 50$. Accordingly, the memory cost of storing the Krylov basis is 
\begin{equation}
M_{\text{Krylov}} = k s d \chi^2.
\end{equation}

During the construction of the Krylov basis, the left and right baths are contracted with the Hamiltonian MPO and the MPS tensors $\psi$. This procedure generates intermediate tensors of size $(\chi, h, d^{2}, \chi)$. The corresponding memory cost is therefore   
\begin{equation}
M_{\text{intermediate}} = 3s h d^2 \chi^2 \approx sd^2\chi^2\sqrt{N}.
\end{equation}

Given that we consider two-level systems ($d = 2$) and use complex 
double-precision floating-point numbers ($s = 16$ bytes) to simulate a 
two-dimensional square lattice with $h_{\mathrm{max}} \approx 3 \sqrt{N}$, 
the final expression for the upper bound of the total memory cost reads 
\begin{equation}
\begin{aligned}
M_{\text{total}} &= M_{\text{MPS}} + M_{\text{baths}} + M_{\text{Krylov}} + M_{\text{intermediate}} \\ 
&\approx \chi^2(s d  N + 3sN^{3/2} + k s d + s d^2 \sqrt{N}) \\
&\approx 48\chi^2 N^{3/2},
\end{aligned}
\end{equation}
where the leading term is fully determined by bath tensors $L_i$ and $R_i$.

\section{Further details of NQS  simulations}
\label{app1}

As a reference to estimate the run-time of NQS simulations, we consider the neural-network architectures used in Refs.~\cite{schmitt2025simulatingdynamicscorrelatedmatter,
Mendes-Santos2023}. These consist of fully connected neural networks (or RBMs) and two-layer CNNs, denoted by $(\alpha_1,\alpha_2;L_f)$, where $\alpha_i$ is related to the number of channels in each layer and $L_f$ is the filter size. In the results of Fig.~\ref{fig:Fig_3}, CNN1 is defined as $(4,3;L/2)$ and CNN2 as $(6,5;L/2)$, while the considered RBM has $\alpha=10$ hidden nodes.

As a figure of merit for the convergence of NQS results, we consider the behavior of the integrated residual $R^2$ (see the main text) for a CNN architecture with $(6,5;L/2)$, shown in Fig.~\ref{fig1_sm}.
The quantity $R^2$ rapidly increases at $t\approx 400\,\mathrm{ns}$ (which corresponds to $tJ\approx 1$), indicating that NQS results struggle to converge to the exact solution.  
As a reference for reliable results, in Fig.~\ref{fig:Fig_5}, we only consider run-time estimates for simulations satisfying $R^2 \leq 0.05$.

\begin{figure}
    \centering
    \includegraphics{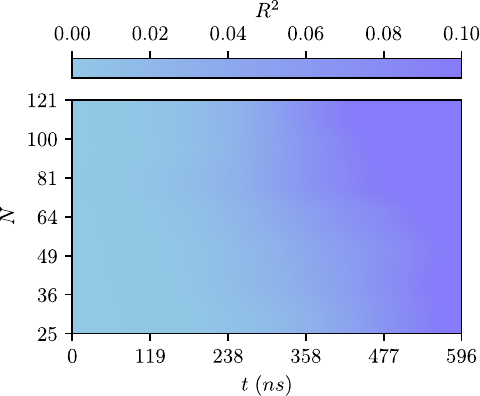}
    \caption{Color plot of the proxy for the TDVP errors in the NQS simulations (see Eq.~\eqref{tdvp_err} in the main text), given by $R^2$, as a function of the total number of qubits $N$ in the square-lattice arrays and the time duration of the dynamics.}
    \label{fig1_sm}
\end{figure}

\begin{figure}
    \centering
    \includegraphics{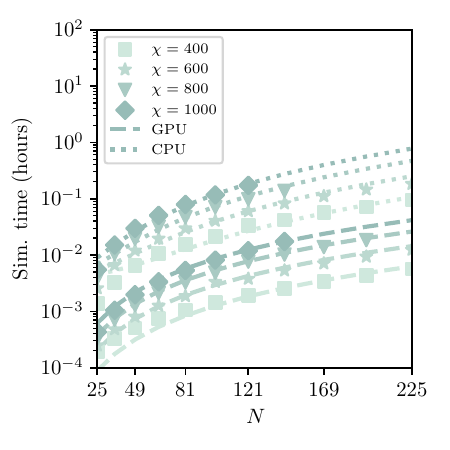}
    \caption{Run-time of a single 1 ns time step after a quantum quench to the Ising model, measured on both CPUs and an NVIDIA A100 GPU, as a function of system size. Results are shown for varying bond dimension and theoretical scaling fits are overlaid.}
    \label{fig:FigSM_1}
\end{figure}

\section{Fitting of GPU run time data}
\label{app:fitting}

In Fig.~\ref{fig:Fig_3}, we presented empirical run-time data measured on an NVIDIA A100 GPU for simulating a single timestep (1 ns) after a quantum quench to the Ising model on square lattices of varying sizes. We considered two classical simulation approaches: MPS, characterized by varying bond dimensions, and NQS, parameterized by the number of variational parameters. For both methods, we included fits based on the theoretical scaling laws derived in the main text. In this section, we describe the fitting procedure in detail.

For the MPS simulations, we collected run-time data across bond dimensions ranging from 100 to 600 and system sizes from $N=25$ to $N= 144$. We then performed a surface fit over this data using a 2D model with just three variational parameters of the form:
\begin{equation}
    t\left(N,\chi\right) = a + bN^{3/2}\chi^3 + c N^2\chi^2,    
\end{equation}
where $N$ is the system size, $\chi$ is the bond dimension, and $a,b,c$ are fit parameters. This fitted surface was then used to extrapolate run-time estimates beyond the original data range, as shown in Fig.~\ref{fig:Fig_3}.

For NQS simulations, we fit a polynomial function of $N$ of degree three:  
$t(N) = a_q N + b_q N^2 + c_q N^3$.
The run-time is expected to be reduced almost linearly with the number of GPUs used in the simulation \cite{Schmitt2022}.  
For the results shown in Fig.~\ref{fig:Fig_3}, in order to keep the amount of resources on an equal footing with MPS, we renormalize the run-time by the number of GPUs in cases where more than one GPU is used.  
This procedure is not applied, however, to the results presented in Fig.~\ref{fig:Fig_4}.

\section{CPU vs GPU}
\label{app:cpu}

In the main text (Fig.~\ref{fig:Fig_3}), we presented simulation results for both MPS and NQS methods, emphasizing the scaling of computational complexity with system size and simulation parameters. Here, we extend this analysis by focusing exclusively on MPS simulations and comparing performance across different hardware backends, see Fig.~\ref{fig:FigSM_1}.

We benchmarked the average run time per time step (with temporal resolution $dt \approx 1$~ns) on both CPU and GPU architectures for identical quench protocols as described in Sec.~\ref{Sec_PC}. The results show that GPU implementations provide a substantial acceleration compared to CPU-based simulations, with typical speedups of approximately one order of magnitude for the system sizes investigated. Importantly, both CPU and GPU data follow the same scaling law, confirming that the underlying complexity analysis remains valid regardless of the hardware platform. The observed difference therefore reflects primarily the efficiency of parallelization and optimized memory access available on GPUs.

This comparison highlights that while classical MPS simulations can in principle scale to moderately large system sizes, the computational resources required depend strongly on the available hardware. In particular, GPU acceleration represents a critical factor for bringing simulations of tens to hundreds of qubits into a tractable regime, and provides a more realistic baseline for evaluating the performance advantage of neutral atom QPUs.

\bibliography{bib}

@article{Mendes-Santos2023,
  title = {Highly Resolved Spectral Functions of Two-Dimensional Systems with Neural Quantum States},
  author = {Mendes-Santos, Tiago and Schmitt, Markus and Heyl, Markus},
  journal = {Phys. Rev. Lett.},
  volume = {131},
  issue = {4},
  pages = {046501},
  numpages = {7},
  year = {2023},
  month = {Jul},
  publisher = {American Physical Society},
  doi = {10.1103/PhysRevLett.131.046501},
  url = {https://link.aps.org/doi/10.1103/PhysRevLett.131.046501}
}

@article{Sinibaldi_2023,
   title={Unbiasing time-dependent Variational Monte Carlo by projected quantum evolution},
   volume={7},
   ISSN={2521-327X},
   url={http://dx.doi.org/10.22331/q-2023-10-10-1131},
   DOI={10.22331/q-2023-10-10-1131},
   journal={Quantum},
   publisher={Verein zur Forderung des Open Access Publizierens in den Quantenwissenschaften},
   author={Sinibaldi, Alessandro and Giuliani, Clemens and Carleo, Giuseppe and Vicentini, Filippo},
   year={2023},
   month=oct, pages={1131} }

@Article{Vicentini2022,
	title={{NetKet 3: Machine Learning Toolbox for Many-Body Quantum Systems}},
	author={Filippo Vicentini and Damian Hofmann and Attila Szabó and Dian Wu and Christopher   Roth and Clemens Giuliani and Gabriel Pescia and Jannes Nys and Vladimir Vargas-Calderón and   Nikita Astrakhantsev and Giuseppe Carleo},
	journal={SciPost Phys. Codebases},
	pages={7},
	year={2022},
	publisher={SciPost},
	doi={10.21468/SciPostPhysCodeb.7},
	url={https://scipost.org/10.21468/SciPostPhysCodeb.7},
}

@Article{Schmitt2022,
	title={{jVMC: Versatile and performant variational Monte Carlo leveraging automated differentiation and GPU acceleration}},
	author={Markus Schmitt and Moritz Reh},
	journal={SciPost Phys. Codebases},
	pages={2},
	year={2022},
	publisher={SciPost},
	doi={10.21468/SciPostPhysCodeb.2},
	url={https://scipost.org/10.21468/SciPostPhysCodeb.2},
}

@article{daley2022practical,
  title={Practical quantum advantage in quantum simulation},
  author={Daley, Andrew J and Bloch, Immanuel and Kokail, Christian and Flannigan, Stuart and Pearson, Natalie and Troyer, Matthias and Zoller, Peter},
  journal={Nature},
  volume={607},
  number={7920},
  pages={667--676},
  year={2022},
  publisher={Nature Publishing Group UK London},
  doi={https://doi.org/10.1038/s41586-022-04940-6}
}

@misc{cai2023stochastic,
  title = {Stochastic Error Cancellation in Analog Quantum Simulation},
  author = {Cai, Yiyi and Tong, Yu and Preskill, John},
  publisher = {Schloss Dagstuhl – Leibniz-Zentrum f\"ur Informatik},
  year = {2024},
  doi = {10.4230/LIPICS.TQC.2024.2},
  url = {https://drops.dagstuhl.de/entities/document/10.4230/LIPIcs.TQC.2024.2},
}

@misc{schiffer2024quantum,
      title={The quantum adiabatic algorithm suppresses the proliferation of errors}, 
      author={Benjamin F. Schiffer and Adrian Franco Rubio and Rahul Trivedi and J. Ignacio Cirac},
      year={2024},
      eprint={2404.15397},
      archivePrefix={arXiv},
      primaryClass={quant-ph},
      url={https://arxiv.org/abs/2404.15397}, 
}

@book{sachdev2023quantum,
  title={Quantum phases of matter},
  author={Sachdev, Subir},
  year={2023},
  publisher={Cambridge University Press},
  doi={https://doi.org/10.1017/9781009212717}
}

@article{pavevsic2025constrained,
  title={Constrained dynamics and confinement in the two-dimensional quantum Ising model},
   volume={111},
   ISSN={2469-9969},
   url={http://dx.doi.org/10.1103/PhysRevB.111.L140305},
   DOI={10.1103/physrevb.111.l140305},
   number={14},
   journal={Physical Review B},
   publisher={American Physical Society (APS)},
   author={Pavešić, Luka and Jaschke, Daniel and Montangero, Simone},
   year={2025},
   month=apr }

@article{krinitsin2025roughening,
  title={Roughening Dynamics of Interfaces in the Two-Dimensional Quantum Ising Model},
   volume={134},
   ISSN={1079-7114},
   url={http://dx.doi.org/10.1103/9bsk-x9rw},
   DOI={10.1103/9bsk-x9rw},
   number={24},
   journal={Physical Review Letters},
   publisher={American Physical Society (APS)},
   author={Krinitsin, Wladislaw and Tausendpfund, Niklas and Rizzi, Matteo and Heyl, Markus and Schmitt, Markus},
   year={2025},
   month=jun }

@article{schmitt_quantum_2022,
	title = {Quantum phase transition dynamics in the two-dimensional transverse-field {Ising} model},
	volume = {8},
	issn = {2375-2548},
	url = {https://www.science.org/doi/10.1126/sciadv.abl6850},
	doi = {10.1126/sciadv.abl6850},
	number = {37},
	urldate = {2023-10-02},
	journal = {Sci. Adv.},
	author = {Schmitt, Markus and Rams, Marek M. and Dziarmaga, Jacek and Heyl, Markus and Zurek, Wojciech H.},
	month = sep,
	year = {2022},
	pages = {eabl6850},
}

@misc{zhang2025probing,
      title={Probing quantum floating phases in {Rydberg} atom arrays}, 
      author={Jin Zhang and Sergio H. Cant\'u and Fangli Liu and Alexei Bylinskii and Boris Braverman and Florian Huber and Jesse Amato-Grill and Alexander Lukin and Nathan Gemelke and Alexander Keesling and others},
      year={2024},
      eprint={2401.08087},
      archivePrefix={arXiv},
      primaryClass={quant-ph},
      url={https://arxiv.org/abs/2401.08087}, 
}

@article{ebadi2021quantum,
   title={Quantum phases of matter on a 256-atom programmable quantum simulator},
   volume={595},
   ISSN={1476-4687},
   url={http://dx.doi.org/10.1038/s41586-021-03582-4},
   DOI={10.1038/s41586-021-03582-4},
   number={7866},
   journal={Nature},
   publisher={Springer Science and Business Media LLC},
   author={Ebadi, Sepehr and Wang, Tout T. and Levine, Harry and Keesling, Alexander and Semeghini, Giulia and Omran, Ahmed and Bluvstein, Dolev and Samajdar, Rhine and Pichler, Hannes and Ho, Wen Wei and others},
   year={2021},
   month=jul, pages={227–232} }

@article{shaw2024benchmarking,
   title={Benchmarking highly entangled states on a 60-atom analogue quantum simulator},
   volume={628},
   ISSN={1476-4687},
   url={http://dx.doi.org/10.1038/s41586-024-07173-x},
   DOI={10.1038/s41586-024-07173-x},
   number={8006},
   journal={Nature},
   publisher={Springer Science and Business Media LLC},
   author={Shaw, Adam L. and Chen, Zhuo and Choi, Joonhee and Mark, Daniel K. and Scholl, Pascal and Finkelstein, Ran and Elben, Andreas and Choi, Soonwon and Endres, Manuel},
   year={2024},
   month=mar, pages={71–77} }

@misc{kornjavca2024large,
      title={Large-scale quantum reservoir learning with an analog quantum computer}, 
      author={Milan Kornjača and Hong-Ye Hu and Chen Zhao and Jonathan Wurtz and Phillip Weinberg and Majd Hamdan and Andrii Zhdanov and Sergio H. Cantu and Hengyun Zhou and Rodrigo Araiza Bravo and others},
      year={2024},
      eprint={2407.02553},
      archivePrefix={arXiv},
      primaryClass={quant-ph},
      url={https://arxiv.org/abs/2407.02553}, 
}

@article{cong2019quantum,
    title={Quantum convolutional neural networks},
   volume={15},
   ISSN={1745-2481},
   url={http://dx.doi.org/10.1038/s41567-019-0648-8},
   DOI={10.1038/s41567-019-0648-8},
   number={12},
   journal={Nature Physics},
   publisher={Springer Science and Business Media LLC},
   author={Cong, Iris and Choi, Soonwon and Lukin, Mikhail D.},
   year={2019},
   month=aug, pages={1273–1278} }

@article{gonzalez2025observation,
   title={Observation of string breaking on a (2+1){D} {Rydberg} quantum simulator},
   volume={642},
   ISSN={1476-4687},
   url={http://dx.doi.org/10.1038/s41586-025-09051-6},
   DOI={10.1038/s41586-025-09051-6},
   number={8067},
   journal={Nature},
   publisher={Springer Science and Business Media LLC},
   author={Gonz\'alez-Cuadra, Daniel and Hamdan, Majd and Zache, Torsten V. and Braverman, Boris and Kornjača, Milan and Lukin, Alexander and Cant\'u, Sergio H. and Liu, Fangli and Wang, Sheng-Tao and Keesling, Alexander and others},
   year={2025},
   month=jun, pages={321–326} }

@article{manovitz2025quantum,
   title={Quantum coarsening and collective dynamics on a programmable simulator},
   volume={638},
   ISSN={1476-4687},
   url={http://dx.doi.org/10.1038/s41586-024-08353-5},
   DOI={10.1038/s41586-024-08353-5},
   number={8049},
   journal={Nature},
   publisher={Springer Science and Business Media LLC},
   author={Manovitz, Tom and Li, Sophie H. and Ebadi, Sepehr and Samajdar, Rhine and Geim, Alexandra A. and Evered, Simon J. and Bluvstein, Dolev and Zhou, Hengyun and Koyluoglu, Nazli Ugur and Feldmeier, Johannes and Dolgirev, Pavel E. and Maskara, Nishad and Kalinowski, Marcin and Sachdev, Subir and Huse, David A. and Greiner, Markus and Vuletić, Vladan and Lukin, Mikhail D.},
   year={2025},
   month=feb, pages={86–92} }

@article{bluvstein2021controlling,
   title={Controlling quantum many-body dynamics in driven {Rydberg} atom arrays},
   volume={371},
   ISSN={1095-9203},
   url={http://dx.doi.org/10.1126/science.abg2530},
   DOI={10.1126/science.abg2530},
   number={6536},
   journal={Science},
   publisher={American Association for the Advancement of Science (AAAS)},
   author={Bluvstein, D. and Omran, A. and Levine, H. and Keesling, A. and Semeghini, G. and Ebadi, S. and Wang, T. T. and Michailidis, A. A. and Maskara, N. and Ho, W. W. and others},
   year={2021},
   month=mar, pages={1355–1359} }

@article{choi2023preparing,
   title={Preparing random states and benchmarking with many-body quantum chaos},
   volume={613},
   ISSN={1476-4687},
   url={http://dx.doi.org/10.1038/s41586-022-05442-1},
   DOI={10.1038/s41586-022-05442-1},
   number={7944},
   journal={Nature},
   publisher={Springer Science and Business Media LLC},
   author={Choi, Joonhee and Shaw, Adam L. and Madjarov, Ivaylo S. and Xie, Xin and Finkelstein, Ran and Covey, Jacob P. and Cotler, Jordan S. and Mark, Daniel K. and Huang, Hsin-Yuan and Kale, Anant and others},
   year={2023},
   month=jan, pages={468–473} }

@misc{bellonzi2024feasibility,
      title={Feasibility of accelerating homogeneous catalyst discovery with fault-tolerant quantum computers}, 
      author={Nicole Bellonzi and Alexander Kunitsa and Joshua T. Cantin and Jorge A. Campos-Gonzalez-Angulo and Maxwell D. Radin and Yanbing Zhou and Peter D. Johnson and Luis A. Mart\'inez-Mart\'inez and Mohammad Reza Jangrouei and Aritra Sankar Brahmachari and others},
      year={2024},
      eprint={2406.06335},
      archivePrefix={arXiv},
      primaryClass={quant-ph},
      url={https://arxiv.org/abs/2406.06335}, 
}

@misc{agrawal2024quantifying,
      title={Quantifying fault tolerant simulation of strongly correlated systems using the Fermi-Hubbard model}, 
      author={Anjali A. Agrawal and Joshua Job and Tyler L. Wilson and S. N. Saadatmand and Mark J. Hodson and Josh Y. Mutus and Athena Caesura and Peter D. Johnson and Justin E. Elenewski and Kaitlyn J. Morrell and Alexander F. Kemper},
      year={2024},
      eprint={2406.06511},
      archivePrefix={arXiv},
      primaryClass={quant-ph},
      url={https://arxiv.org/abs/2406.06511}, 
}

@misc{nguyen2024quantum,
  title={Quantum computing for corrosion-resistant materials and anti-corrosive coatings design}, 
      author={Nam Nguyen and Thomas W. Watts and Benjamin Link and Kristen S. Williams and Yuval R. Sanders and Samuel J. Elman and Maria Kieferova and Michael J. Bremner and Kaitlyn J. Morrell and Justin Elenewski and others},
      year={2025},
      eprint={2406.18759},
      archivePrefix={arXiv},
      primaryClass={quant-ph},
      url={https://arxiv.org/abs/2406.18759}, 
}

@misc{huang2025vast,
      title={The vast world of quantum advantage}, 
      author={Hsin-Yuan Huang and Soonwon Choi and Jarrod R. McClean and John Preskill},
      year={2025},
      eprint={2508.05720},
      archivePrefix={arXiv},
      primaryClass={quant-ph},
      url={https://arxiv.org/abs/2508.05720}, 
}

@article{carleo_solving_2017,
	title = {Solving the {Quantum} {Many}-{Body} {Problem} with {Artificial} {Neural} {Networks}},
	volume = {355},
	issn = {0036-8075, 1095-9203},
	url = {http://arxiv.org/abs/1606.02318},
	doi = {10.1126/science.aag2302},
	number = {6325},
	urldate = {2024-08-26},
	journal = {Science},
	author = {Carleo, Giuseppe and Troyer, Matthias},
	month = feb,
	year = {2017},
	pages = {602--606},
}

@misc{dalyac2024graph,
      title={Graph Algorithms with Neutral Atom Quantum Processors}, 
      author={Constantin Dalyac and Lucas Leclerc and Louis Vignoli and Mehdi Djellabi and Wesley da Silva Coelho and Bruno Ximenez and Alexandre Dareau and Davide Dreon and Vincent E. Elfving and Adrien Signoles and others},
      year={2024},
      eprint={2403.11931},
      archivePrefix={arXiv},
      primaryClass={quant-ph},
      url={https://arxiv.org/abs/2403.11931}, 
}

@article{suarez2025energy,
   title={Energy efficiency trends in HPC: what high-energy and astrophysicists need to know},
   volume={13},
   ISSN={2296-424X},
   url={http://dx.doi.org/10.3389/fphy.2025.1542474},
   DOI={10.3389/fphy.2025.1542474},
   journal={Frontiers in Physics},
   publisher={Frontiers Media SA},
   author={Suarez, Estela and Amaya, Jorge and Frank, Martin and Freyermuth, Oliver and Girone, Maria and Kostrzewa, Bartosz and Pfalzner, Susanne},
   year={2025},
   month=apr }

@misc{lanes2025frameworkquantumadvantage,
      title={A Framework for Quantum Advantage}, 
      author={Olivia Lanes and Mourad Beji and Antonio D. Corcoles and Constantin Dalyac and Jay M. Gambetta and Loic Henriet and Ali Javadi-Abhari and Abhinav Kandala and Antonio Mezzacapo and Christopher Porter and others},
      year={2025},
      eprint={2506.20658},
      archivePrefix={arXiv},
      primaryClass={quant-ph},
      url={https://arxiv.org/abs/2506.20658}, 
}

@book{nielsen2010quantum,
  title={Quantum computation and quantum information},
  author={Nielsen, Michael A and Chuang, Isaac L},
  year={2010},
  publisher={Cambridge university press},
  doi = {https://doi.org/10.1017/CBO9780511976667}
}

@article{Henriet_2020,
   title={Quantum computing with neutral atoms},
   volume={4},
   ISSN={2521-327X},
   url={http://dx.doi.org/10.22331/q-2020-09-21-327},
   DOI={10.22331/q-2020-09-21-327},
   journal={Quantum},
   publisher={Verein zur Forderung des Open Access Publizierens in den Quantenwissenschaften},
   author={Henriet, Lo\"ic and Beguin, Lucas and Signoles, Adrien and Lahaye, Thierry and Browaeys, Antoine and Reymond, Georges-Olivier and Jurczak, Christophe},
   year={2020},
   month=sep, pages={327} }

@article{Browaeys_2020,
   title={Many-body physics with individually controlled {Rydberg} atoms},
   volume={16},
   ISSN={1745-2481},
   url={http://dx.doi.org/10.1038/s41567-019-0733-z},
   DOI={10.1038/s41567-019-0733-z},
   number={2},
   journal={Nature Physics},
   publisher={Springer Science and Business Media LLC},
   author={Browaeys, Antoine and Lahaye, Thierry},
   year={2020},
   month=jan, pages={132–142} }

@article{Scholl_2021,
   title={Quantum simulation of {2D} antiferromagnets with hundreds of {Rydberg} atoms},
   volume={595},
   ISSN={1476-4687},
   url={http://dx.doi.org/10.1038/s41586-021-03585-1},
   DOI={10.1038/s41586-021-03585-1},
   number={7866},
   journal={Nature},
   publisher={Springer Science and Business Media LLC},
   author={Scholl, Pascal and Schuler, Michael and Williams, Hannah J. and Eberharter, Alexander A. and Barredo, Daniel and Schymik, Kai-Niklas and Lienhard, Vincent and Henry, Louis-Paul and Lang, Thomas C. and Lahaye, Thierry and others},
   year={2021},
   month=jul, pages={233–238} }

@article{Ebadi_2022,
   title={Quantum optimization of maximum independent set using {Rydberg} atom arrays},
   volume={376},
   ISSN={1095-9203},
   url={http://dx.doi.org/10.1126/science.abo6587},
   DOI={10.1126/science.abo6587},
   number={6598},
   journal={Science},
   publisher={American Association for the Advancement of Science (AAAS)},
   author={Ebadi, S. and Keesling, A. and Cain, M. and Wang, T. T. and Levine, H. and Bluvstein, D. and Semeghini, G. and Omran, A. and Liu, J.-G. and Samajdar, R. and others},
   year={2022},
   month=jun, pages={1209–1215} }

@article{Semeghini_2021,
   title={Probing topological spin liquids on a programmable quantum simulator},
   volume={374},
   ISSN={1095-9203},
   url={http://dx.doi.org/10.1126/science.abi8794},
   DOI={10.1126/science.abi8794},
   number={6572},
   journal={Science},
   publisher={American Association for the Advancement of Science (AAAS)},
   author={Semeghini, G. and Levine, H. and Keesling, A. and Ebadi, S. and Wang, T. T. and Bluvstein, D. and Verresen, R. and Pichler, H. and Kalinowski, M. and Samajdar, R. and others},
   year={2021},
   month=dec, pages={1242–1247} }

@misc{vovrosh2025mesonspectroscopyexoticsymmetries,
      title={Meson spectroscopy of exotic symmetries of Ising criticality in {Rydberg} atom arrays}, 
      author={Joseph Vovrosh and Julius de Hond and Sergi Juli\a`-Farr\'e and Johannes Knolle and Alexandre Dauphin},
      year={2025},
      eprint={2506.21299},
      archivePrefix={arXiv},
      primaryClass={quant-ph},
      url={https://arxiv.org/abs/2506.21299}, 
}

@article{henry2021quantum,
   title={Quantum evolution kernel: Machine learning on graphs with programmable arrays of qubits},
   volume={104},
   ISSN={2469-9934},
   url={http://dx.doi.org/10.1103/PhysRevA.104.032416},
   DOI={10.1103/physreva.104.032416},
   number={3},
   journal={Physical Review A},
   publisher={American Physical Society (APS)},
   author={Henry, Louis-Paul and Thabet, Slimane and Dalyac, Constantin and Henriet, Lo\"ic},
   year={2021},
   month=sep }

@misc{bernstein1993quantum,
  title={Quantum complexity theory},
  author={Bernstein, Ethan and Vazirani, Umesh},
  booktitle={Proceedings of the twenty-fifth annual ACM symposium on Theory of computing},
  pages={11--20},
  year={1993},
  doi = {https://dl.acm.org/doi/10.1145/167088.167097}
}

@article{jaschke2023quantum,
   title={Is quantum computing green? An estimate for an energy-efficiency quantum advantage},
   volume={8},
   ISSN={2058-9565},
   number={2},
   journal={Quantum Science and Technology},
   publisher={IOP Publishing},
   author={Jaschke, Daniel and Montangero, Simone},
   year={2023},
   month=jan, 
   pages={025001},
   url={http://dx.doi.org/10.1088/2058-9565/acae3e},
   doi={10.1088/2058-9565/acae3e},
}

@article{fellous2021limitations,
  title = {Limitations in Quantum Computing from Resource Constraints},
  author = {Fellous-Asiani, Marco and Chai, Jing Hao and Whitney, Robert S. and Auff\`eves, Alexia and Ng, Hui Khoon},
  journal = {PRX Quantum},
  volume = {2},
  issue = {4},
  pages = {040335},
  numpages = {11},
  year = {2021},
  month = {Nov},
  publisher = {American Physical Society},
  doi = {10.1103/PRXQuantum.2.040335},
  url = {https://link.aps.org/doi/10.1103/PRXQuantum.2.040335}
}

@misc{van2023using,
  title={Using Azure Quantum Resource Estimator for Assessing Performance of Fault Tolerant Quantum Computation}, 
      author={Wim van Dam and Mariia Mykhailova and Mathias Soeken},
      year={2024},
      eprint={2311.05801},
      archivePrefix={arXiv},
      primaryClass={quant-ph},
      url={https://arxiv.org/abs/2311.05801}, 
}

@misc{gidney2025factor,
  title={How to factor 2048 bit {RSA} integers with less than a million noisy qubits}, 
      author={Craig Gidney},
      year={2025},
      eprint={2505.15917},
      archivePrefix={arXiv},
      primaryClass={quant-ph},
      url={https://arxiv.org/abs/2505.15917}, 
}

@misc{harrigan2024expressing,
  title={Expressing and Analyzing Quantum Algorithms with Qualtran}, 
      author={Matthew P. Harrigan and Tanuj Khattar and Charles Yuan and Anurudh Peduri and Noureldin Yosri and Fionn D. Malone and Ryan Babbush and Nicholas C. Rubin},
      year={2024},
      eprint={2409.04643},
      archivePrefix={arXiv},
      primaryClass={quant-ph},
      url={https://arxiv.org/abs/2409.04643}, 
}

@misc{zhou2025resourceanalysislowoverheadtransversal,
      title={Resource Analysis of Low-Overhead Transversal Architectures for Reconfigurable Atom Arrays}, 
      author={Hengyun Zhou and Casey Duckering and Chen Zhao and Dolev Bluvstein and Madelyn Cain and Aleksander Kubica and Sheng-Tao Wang and Mikhail D. Lukin},
      year={2025},
      eprint={2505.15907},
      archivePrefix={arXiv},
      primaryClass={quant-ph},
      url={https://arxiv.org/abs/2505.15907}, 
}

@article{auffeves2022quantum,
  title = {Quantum Technologies Need a Quantum Energy Initiative},
  author = {Auff\`eves, Alexia},
  journal = {PRX Quantum},
  volume = {3},
  issue = {2},
  pages = {020101},
  numpages = {12},
  year = {2022},
  month = {Jun},
  publisher = {American Physical Society},
  doi = {10.1103/PRXQuantum.3.020101},
  url = {https://link.aps.org/doi/10.1103/PRXQuantum.3.020101}
}

@article{Bernien_2017,
   title={Probing many-body dynamics on a 51-atom quantum simulator},
   volume={551},
   ISSN={1476-4687},
   url={http://dx.doi.org/10.1038/nature24622},
   DOI={10.1038/nature24622},
   number={7682},
   journal={Nature},
   publisher={Springer Science and Business Media LLC},
   author={Bernien, Hannes and Schwartz, Sylvain and Keesling, Alexander and Levine, Harry and Omran, Ahmed and Pichler, Hannes and Choi, Soonwon and Zibrov, Alexander S. and Endres, Manuel and Greiner, Markus and others},
   year={2017},
   month=nov, pages={579–584} }

@misc{cazals2025identifyinghardnativeinstances,
      title={Identifying hard native instances for the maximum independent set problem on neutral atoms quantum processors}, 
      author={Pierre Cazals and Aymeric Fran\c{c}ois and Lo\"ic Henriet and Lucas Leclerc and Malory Marin and Yassine Naghmouchi and Wesley da Silva Coelho and Florian Sikora and Vittorio Vitale and R\'emi Watrigant and others},
      year={2025},
      eprint={2502.04291},
      archivePrefix={arXiv},
      primaryClass={quant-ph},
      url={https://arxiv.org/abs/2502.04291}, 
}

@article{hauschild2018efficient,
  title={Efficient numerical simulations with tensor networks: Tensor Network Python (TeNPy)},
  author={Hauschild, Johannes and Pollmann, Frank},
  journal={SciPost Physics Lecture Notes},
  pages={005},
  year={2018},
 doi = {doi: 10.21468/SciPostPhysLectNotes.5}
}

@article{fishman2022itensor,
  title={The ITensor software library for tensor network calculations},
  author={Fishman, Matthew and White, Steven and Stoudenmire, Edwin Miles},
  journal={SciPost Physics Codebases},
  pages={004},
  year={2022},
doi = {doi: 10.21468/SciPostPhysCodeb.4}
}

@misc{schmitt2025simulatingdynamicscorrelatedmatter,
      title={Simulating dynamics of correlated matter with neural quantum states}, 
      author={Markus Schmitt and Markus Heyl},
      year={2025},
      eprint={2506.03124},
      archivePrefix={arXiv},
      primaryClass={quant-ph},
      url={https://arxiv.org/abs/2506.03124}, 
}

@misc{emu-mps,
  doi = {10.48550/ARXIV.2510.09813},
  url = {https://arxiv.org/abs/2510.09813},
  author = {Bidzhiev,  Kemal and Grava,  Stefano and Henaff,  Pablo le and Mendizabal,  Mauro and Merhej,  Elie and Quelle,  Anton},
  title = {Efficient Emulation of Neutral Atom Quantum Hardware},
  publisher = {arXiv},
  year = {2025},
  copyright = {arXiv.org perpetual,  non-exclusive license}
}

@article{Haegeman_tdvp_1,
  title = {Time-Dependent Variational Principle for Quantum Lattices},
  author = {Haegeman, Jutho and Cirac, J. Ignacio and Osborne, Tobias J. and Pi\ifmmode \check{z}\else \v{z}\fi{}orn, Iztok and Verschelde, Henri and Verstraete, Frank},
  journal = {Phys. Rev. Lett.},
  volume = {107},
  issue = {7},
  pages = {070601},
  numpages = {5},
  year = {2011},
  month = {Aug},
  publisher = {American Physical Society},
  doi = {10.1103/PhysRevLett.107.070601},
  url = {https://link.aps.org/doi/10.1103/PhysRevLett.107.070601}
}

@article{Haegeman_tdvp_2,
  title = {Unifying time evolution and optimization with matrix product states},
  author = {Haegeman, Jutho and Lubich, Christian and Oseledets, Ivan and Vandereycken, Bart and Verstraete, Frank},
  journal = {Phys. Rev. B},
  volume = {94},
  issue = {16},
  pages = {165116},
  numpages = {10},
  year = {2016},
  month = {Oct},
  publisher = {American Physical Society},
  doi = {10.1103/PhysRevB.94.165116},
  url = {https://link.aps.org/doi/10.1103/PhysRevB.94.165116}
}

@article{paeckel_mps,
title = {Time-evolution methods for matrix-product states},
journal = {Annals of Physics},
volume = {411},
pages = {167998},
year = {2019},
issn = {0003-4916},
doi = {https://doi.org/10.1016/j.aop.2019.167998},
url = {https://www.sciencedirect.com/science/article/pii/S0003491619302532},
author = {Sebastian Paeckel and Thomas K\"ohler and Andreas Swoboda and Salvatore R. Manmana and Ulrich Schollw\"ock and Claudius Hubig},}

@article{crosswhite_mpo,
  title = {Applying matrix product operators to model systems with long-range interactions},
  author = {Crosswhite, Gregory M. and Doherty, A. C. and Vidal, Guifr\'e},
  journal = {Phys. Rev. B},
  volume = {78},
  issue = {3},
  pages = {035116},
  numpages = {7},
  year = {2008},
  month = {Jul},
  publisher = {American Physical Society},
  doi = {10.1103/PhysRevB.78.035116},
  url = {https://link.aps.org/doi/10.1103/PhysRevB.78.035116}
}

@article{lancz_matrix_exp,
  title = {On Krylov Subspace Approximations to the Matrix Exponential Operator},
  volume = {34},
  ISSN = {1095-7170},
  url = {http://dx.doi.org/10.1137/S0036142995280572},
  DOI = {10.1137/s0036142995280572},
  number = {5},
  journal = {SIAM Journal on Numerical Analysis},
  publisher = {Society for Industrial & Applied Mathematics (SIAM)},
  author = {Hochbruck,  Marlis and Lubich,  Christian},
  year = {1997},
  month = oct,
  pages = {1911–1925}
}

@article{Lee_2017,
   title={Defect-free atomic array formation using the Hungarian matching algorithm},
   volume={95},
   ISSN={2469-9934},
   url={http://dx.doi.org/10.1103/PhysRevA.95.053424},
   DOI={10.1103/physreva.95.053424},
   number={5},
   journal={Physical Review A},
   publisher={American Physical Society (APS)},
   author={Lee, Woojun and Kim, Hyosub and Ahn, Jaewook},
   year={2017},
   month=may }

@article{Barredo_2016,
   title={An atom-by-atom assembler of defect-free arbitrary two-dimensional atomic arrays},
   volume={354},
   ISSN={1095-9203},
   url={http://dx.doi.org/10.1126/science.aah3778},
   DOI={10.1126/science.aah3778},
   number={6315},
   journal={Science},
   publisher={American Association for the Advancement of Science (AAAS)},
   author={Barredo, Daniel and de L\'es\'eleuc, Sylvain and Lienhard, Vincent and Lahaye, Thierry and Browaeys, Antoine},
   year={2016},
   month=nov, pages={1021–1023} }

@misc{lin2024aienabledrapidassemblythousands,
      title={{AI}-Enabled Rapid Assembly of Thousands of Defect-Free Neutral Atom Arrays with Constant-time-overhead}, 
      author={Rui Lin and Han-Sen Zhong and You Li and Zhang-Rui Zhao and Le-Tian Zheng and Tai-Ran Hu and Hong-Ming Wu and Zhan Wu and Wei-Jie Ma and Yan Gao and others},
      year={2024},
      eprint={2412.14647},
      archivePrefix={arXiv},
      primaryClass={quant-ph},
      url={https://arxiv.org/abs/2412.14647}, 
}

@article{Schymik_2020,
   title={Enhanced atom-by-atom assembly of arbitrary tweezer arrays},
   volume={102},
   ISSN={2469-9934},
   url={http://dx.doi.org/10.1103/PhysRevA.102.063107},
   DOI={10.1103/physreva.102.063107},
   number={6},
   journal={Physical Review A},
   publisher={American Physical Society (APS)},
   author={Schymik, Kai-Niklas and Lienhard, Vincent and Barredo, Daniel and Scholl, Pascal and Williams, Hannah and Browaeys, Antoine and Lahaye, Thierry},
   year={2020},
   month=dec }

@misc{schwartz2019greenai,
      title={Green {AI}}, 
      author={Roy Schwartz and Jesse Dodge and Noah A. Smith and Oren Etzioni},
      year={2019},
      eprint={1907.10597},
      archivePrefix={arXiv},
      primaryClass={cs.CY},
      url={https://arxiv.org/abs/1907.10597}, 
}

@misc{verdecchia2023systematicreviewgreenai,
      title={A Systematic Review of Green {AI}}, 
      author={Roberto Verdecchia and June Sallou and Lu\'is Cruz},
      year={2023},
      eprint={2301.11047},
      archivePrefix={arXiv},
      primaryClass={cs.AI},
      url={https://arxiv.org/abs/2301.11047}, 
}

@article{PhysRevA.108.023107,
  title = {Efficient algorithms to solve atom reconfiguration problems. {I}. Redistribution-reconfiguration algorithm},
  author = {Cimring, Barry and El Sabeh, Remy and Bacvanski, Marc and Maaz, Stephanie and El Hajj, Izzat and Nishimura, Naomi and Mouawad, Amer E. and Cooper, Alexandre},
  journal = {Phys. Rev. A},
  volume = {108},
  issue = {2},
  pages = {023107},
  numpages = {14},
  year = {2023},
  month = {Aug},
  publisher = {American Physical Society},
  doi = {10.1103/PhysRevA.108.023107},
  url = {https://link.aps.org/doi/10.1103/PhysRevA.108.023107}
}

@article{PhysRevA.108.023108,
  title = {Efficient algorithms to solve atom reconfiguration problems. {II}. Assignment-rerouting-ordering algorithm},
  author = {El Sabeh, Remy and Bohm, Jessica and Ding, Zhiqian and Maaz, Stephanie and Nishimura, Naomi and El Hajj, Izzat and Mouawad, Amer E. and Cooper, Alexandre},
  journal = {Phys. Rev. A},
  volume = {108},
  issue = {2},
  pages = {023108},
  numpages = {24},
  year = {2023},
  month = {Aug},
  publisher = {American Physical Society},
  doi = {10.1103/PhysRevA.108.023108},
  url = {https://link.aps.org/doi/10.1103/PhysRevA.108.023108}
}

@article{arute2019quantum,
  author       = {Frank Arute and Kunal Arya and Ryan Babbush and Dave Bacon and Joseph C. Bardin and Rami Barends and Rupak Biswas and Sergio Boixo and Fernando G. S. L. Brandão and David A. Buell and others},
  title        = {Quantum supremacy using a programmable superconducting processor},
  journal      = {Nature},
  year         = {2019},
  volume       = {574},
  pages        = {505--510},
  doi          = {10.1038/s41586-019-1666-5}
}

@misc{góis2024energeticquantumadvantagetrappedion,
      title={Towards Energetic Quantum Advantage in Trapped-Ion Quantum Computation}, 
      author={Francisca Góis and Marco Pezzutto and Yasser Omar},
      year={2024},
      eprint={2404.11572},
      archivePrefix={arXiv},
      primaryClass={quant-ph},
      url={https://arxiv.org/abs/2404.11572}, 
}

@article{cimini2020experimental,
  title={Experimental characterization of the energetics of quantum logic gates},
  author={Cimini, Valeria and Gherardini, Stefano and Barbieri, Marco and Gianani, Ilaria and Sbroscia, Marco and Buffoni, Lorenzo and Paternostro, Mauro and Caruso, Filippo},
  journal={npj Quantum Information},
  volume={6},
  number={1},
  pages={96},
  year={2020},
  publisher={Nature Publishing Group UK London},
  doi={https://doi.org/10.1038/s41534-020-00325-7}
}

@article{2rds-k4n9,
  title = {Efficient algorithms to solve atom reconfiguration problems. {III}. The bird and batching algorithms and other parallel implementations on GPUs},
  author = {Afiouni, Fouad and El Sabeh, Remy and Nishimura, Naomi and El Hajj, Izzat and Mouawad, Amer E. and Cooper, Alexandre},
  journal = {Phys. Rev. A},
  volume = {112},
  issue = {2},
  pages = {023109},
  numpages = {17},
  year = {2025},
  month = {Aug},
  publisher = {American Physical Society},
  doi = {10.1103/2rds-k4n9},
  url = {https://link.aps.org/doi/10.1103/2rds-k4n9}
}

@misc{li2025fastcontinuouscoherentatom,
      title={Fast, continuous and coherent atom replacement in a neutral atom qubit array}, 
      author={Yiyi Li and Yicheng Bao and Michael Peper and Chenyuan Li and Jeff D. Thompson},
      year={2025},
      eprint={2506.15633},
      archivePrefix={arXiv},
      primaryClass={quant-ph},
      url={https://arxiv.org/abs/2506.15633}, 
}

@misc{chiu2025continuousoperationcoherent3000qubit,
      title={Continuous operation of a coherent 3,000-qubit system}, 
      author={Neng-Chun Chiu and Elias C. Trapp and Jinen Guo and Mohamed H. Abobeih and Luke M. Stewart and Simon Hollerith and Pavel Stroganov and Marcin Kalinowski and Alexandra A. Geim and Simon J. Evered and others},
      year={2025},
      eprint={2506.20660},
      archivePrefix={arXiv},
      primaryClass={quant-ph},
      url={https://arxiv.org/abs/2506.20660}, 
}

@article{Morvan_2024,
   title={Phase transitions in random circuit sampling},
   volume={634},
   ISSN={1476-4687},
   url={http://dx.doi.org/10.1038/s41586-024-07998-6},
   DOI={10.1038/s41586-024-07998-6},
   number={8033},
   journal={Nature},
   publisher={Springer Science and Business Media LLC},
   author={Morvan, A. and Villalonga, B. and Mi, X. and Mandr\a`, S. and Bengtsson, A. and Klimov, P. V. and Chen, Z. and Hong, S. and Erickson, C. and Drozdov, I. K. and Chau, J. and Laun, G. and others},
   year={2024},
   month=oct, pages={328–333} }

@misc{haghshenas2025digitalquantummagnetismfrontier,
      title={Digital quantum magnetism at the frontier of classical simulations}, 
      author={Reza Haghshenas and Eli Chertkov and Michael Mills and Wilhelm Kadow and Sheng-Hsuan Lin and Yi-Hsiang Chen and Chris Cade and Ido Niesen and Tomislav Begušić and Manuel S. Rudolph and others},
      year={2025},
      eprint={2503.20870},
      archivePrefix={arXiv},
      primaryClass={quant-ph},
      url={https://arxiv.org/abs/2503.20870}, 
}

@misc{liu2025robustquantumcomputationaladvantage,
      title={Robust quantum computational advantage with programmable 3050-photon Gaussian boson sampling}, 
      author={Hua-Liang Liu and Hao Su and Si-Qiu Gong and Yi-Chao Gu and Hao-Yang Tang and Meng-Hao Jia and Qian Wei and Yukun Song and Dongzhou Wang and Mingyang Zheng and others},
      year={2025},
      eprint={2508.09092},
      archivePrefix={arXiv},
      primaryClass={quant-ph},
      url={https://arxiv.org/abs/2508.09092}, 
}

@misc{nvidia_a100_datasheet,
  author       = {NVIDIA Corporation},
  title        = {NVIDIA {A100} Tensor Core GPU Datasheet},
  year         = 2021,
  url          = {https://www.nvidia.com/content/dam/en-zz/Solutions/Data-Center/a100/pdf/nvidia-a100-datasheet-us-nvidia-1758950-r4-web.pdf},
  note         = {Accessed: 2025-09-16}
}

@article{Paeckel_2019,
   title={Time-evolution methods for matrix-product states},
   volume={411},
   ISSN={0003-4916},
   url={http://dx.doi.org/10.1016/j.aop.2019.167998},
   DOI={10.1016/j.aop.2019.167998},
   journal={Annals of Physics},
   publisher={Elsevier BV},
   author={Paeckel, Sebastian and Köhler, Thomas and Swoboda, Andreas and Manmana, Salvatore R. and Schollwöck, Ulrich and Hubig, Claudius},
   year={2019},
   month=dec, pages={167998} }

@article{Wu_2022,
   title={Erasure conversion for fault-tolerant quantum computing in alkaline earth Rydberg atom arrays},
   volume={13},
   ISSN={2041-1723},
   url={http://dx.doi.org/10.1038/s41467-022-32094-6},
   DOI={10.1038/s41467-022-32094-6},
   number={1},
   journal={Nature Communications},
   publisher={Springer Science and Business Media LLC},
   author={Wu, Yue and Kolkowitz, Shimon and Puri, Shruti and Thompson, Jeff D.},
   year={2022},
   month=aug }

@misc{holland2024demonstrationerasureconversionmolecular,
      title={Demonstration of Erasure Conversion in a Molecular Tweezer Array}, 
      author={Connor M. Holland and Yukai Lu and Samuel J. Li and Callum L. Welsh and Lawrence W. Cheuk},
      year={2024},
      eprint={2406.02391},
      archivePrefix={arXiv},
      primaryClass={quant-ph},
      url={https://arxiv.org/abs/2406.02391}, 
}

@misc{vovrosh2025simulatingdynamicstwodimensionaltransversefield,
      title={Simulating dynamics of the two-dimensional transverse-field Ising model: a comparative study of large-scale classical numerics}, 
      author={Joseph Vovrosh and Sergi Julià-Farré and Wladislaw Krinitsin and Michael Kaicher and Fergus Hayes and Emmanuel Gottlob and Augustine Kshetrimayum and Kemal Bidzhiev and Simon B. Jäger and Markus Schmitt and others},
      year={2025},
      eprint={2511.19340},
      archivePrefix={arXiv},
      primaryClass={quant-ph},
      url={https://arxiv.org/abs/2511.19340}, 
}

@misc{krinitsin2025timeevolutionquantumising,
      title={Time evolution of the quantum Ising model in two dimensions using Tree Tensor Networks}, 
      author={Wladislaw Krinitsin and Niklas Tausendpfund and Markus Heyl and Matteo Rizzi and Markus Schmitt},
      year={2025},
      eprint={2505.07612},
      archivePrefix={arXiv},
      primaryClass={quant-ph},
      url={https://arxiv.org/abs/2505.07612}, 
}

@article{Krinitsin_2025,
   title={Roughening Dynamics of Interfaces in the Two-Dimensional Quantum Ising Model},
   volume={134},
   ISSN={1079-7114},
   url={http://dx.doi.org/10.1103/9bsk-x9rw},
   DOI={10.1103/9bsk-x9rw},
   number={24},
   journal={Physical Review Letters},
   publisher={American Physical Society (APS)},
   author={Krinitsin, Wladislaw and Tausendpfund, Niklas and Rizzi, Matteo and Heyl, Markus and Schmitt, Markus},
   year={2025},
   month=jun }

@article{Pave_i__2025,
   title={Constrained dynamics and confinement in the two-dimensional quantum Ising model},
   volume={111},
   ISSN={2469-9969},
   url={http://dx.doi.org/10.1103/PhysRevB.111.L140305},
   DOI={10.1103/physrevb.111.l140305},
   number={14},
   journal={Physical Review B},
   publisher={American Physical Society (APS)},
   author={Pavešić, Luka and Jaschke, Daniel and Montangero, Simone},
   year={2025},
   month=apr }

@misc{pavešić2025scatteringinducedfalsevacuum,
      title={Scattering and induced false vacuum decay in the two-dimensional quantum Ising model}, 
      author={Luka Pavešić and Marco Di Liberto and Simone Montangero},
      year={2025},
      eprint={2509.02702},
      archivePrefix={arXiv},
      primaryClass={quant-ph},
      url={https://arxiv.org/abs/2509.02702}, 
}

@misc{gibson2025quantumcountingrydbergblockade,
      title={Quantum Counting in the Rydberg Blockade}, 
      author={Joseph Gibson and Victor Drouin-Touchette and Stefanos Kourtis},
      year={2025},
      eprint={2506.19298},
      archivePrefix={arXiv},
      primaryClass={quant-ph},
      url={https://arxiv.org/abs/2506.19298}, 
}

@article{PRXQuantum.5.010308,
  title = {Efficient Tensor Network Simulation of IBM's Eagle Kicked Ising Experiment},
  author = {Tindall, Joseph and Fishman, Matthew and Stoudenmire, E. Miles and Sels, Dries},
  journal = {PRX Quantum},
  volume = {5},
  issue = {1},
  pages = {010308},
  numpages = {16},
  year = {2024},
  month = {Jan},
  publisher = {American Physical Society},
  doi = {10.1103/PRXQuantum.5.010308},
  url = {https://link.aps.org/doi/10.1103/PRXQuantum.5.010308}
}

@misc{tindall2025dynamicsdisorderedquantumsystems,
      title={Dynamics of disordered quantum systems with two- and three-dimensional tensor networks}, 
      author={Joseph Tindall and Antonio Mello and Matt Fishman and Miles Stoudenmire and Dries Sels},
      year={2025},
      eprint={2503.05693},
      archivePrefix={arXiv},
      primaryClass={quant-ph},
      url={https://arxiv.org/abs/2503.05693}, 
}

@misc{rudolph2025simulatingsamplingquantumcircuits,
      title={Simulating and Sampling from Quantum Circuits with 2D Tensor Networks}, 
      author={Manuel S. Rudolph and Joseph Tindall},
      year={2025},
      eprint={2507.11424},
      archivePrefix={arXiv},
      primaryClass={quant-ph},
      url={https://arxiv.org/abs/2507.11424}, 
}

@article{Begu_i__2024,
   title={Fast and converged classical simulations of evidence for the utility of quantum computing before fault tolerance},
   volume={10},
   ISSN={2375-2548},
   url={http://dx.doi.org/10.1126/sciadv.adk4321},
   DOI={10.1126/sciadv.adk4321},
   number={3},
   journal={Science Advances},
   publisher={American Association for the Advancement of Science (AAAS)},
   author={Begušić, Tomislav and Gray, Johnnie and Chan, Garnet Kin-Lic},
   year={2024},
   month=jan }

@article{7jzt-xhn6,
  title = {Simulating quantum dynamics in two-dimensional lattices with tensor network influence functional belief propagation},
  author = {Park, Gunhee and Gray, Johnnie and Chan, Garnet Kin-Lic},
  journal = {Phys. Rev. B},
  volume = {112},
  issue = {17},
  pages = {174310},
  numpages = {16},
  year = {2025},
  month = {Nov},
  publisher = {American Physical Society},
  doi = {10.1103/7jzt-xhn6},
  url = {https://link.aps.org/doi/10.1103/7jzt-xhn6}
}

@misc{broers2024exclusiveorencodedalgebraicstructure,
      title={Exclusive-or encoded algebraic structure for efficient quantum dynamics}, 
      author={Lukas Broers and Ludwig Mathey},
      year={2024},
      eprint={2404.09312},
      archivePrefix={arXiv},
      primaryClass={cond-mat.other},
      url={https://arxiv.org/abs/2404.09312}, 
}

@misc{broers2025scalablesimulationquantummanybody,
      title={Scalable Simulation of Quantum Many-Body Dynamics with Or-Represented Quantum Algebra}, 
      author={Lukas Broers and Rong-Yang Sun and Seiji Yunoki},
      year={2025},
      eprint={2506.13241},
      archivePrefix={arXiv},
      primaryClass={quant-ph},
      url={https://arxiv.org/abs/2506.13241}, 
}

@article{PRXQuantum.6.020302,
  title = {Real-Time Operator Evolution in Two and Three Dimensions via Sparse Pauli Dynamics},
  author = {Begu\ifmmode \check{s}\else \v{s}\fi{}i\ifmmode \acute{c}\else \'{c}\fi{}, Tomislav and Chan, Garnet Kin-Lic},
  journal = {PRX Quantum},
  volume = {6},
  issue = {2},
  pages = {020302},
  numpages = {13},
  year = {2025},
  month = {Apr},
  publisher = {American Physical Society},
  doi = {10.1103/PRXQuantum.6.020302},
  url = {https://link.aps.org/doi/10.1103/PRXQuantum.6.020302}
}

@misc{rudolph2025paulipropagationcomputationalframework,
      title={Pauli Propagation: A Computational Framework for Simulating Quantum Systems}, 
      author={Manuel S. Rudolph and Tyson Jones and Yanting Teng and Armando Angrisani and Zoë Holmes},
      year={2025},
      eprint={2505.21606},
      archivePrefix={arXiv},
      primaryClass={quant-ph},
      url={https://arxiv.org/abs/2505.21606}, 
}

\end{document}